\overfullrule 0pt
\nopagenumbers
\font\big=cmbx12 at 14pt
\font\bigit=cmti12 at 14pt
\input colordvi
\def\h{\hfill\break}\nopagenumbers
\parindent=0pt
\magnification\magstep2
\voffset=-5truemm
\hoffset -5truemm
\hsize=170truemm\vsize=270truemm
\def\b{\BrickRed{$\bullet ~$}}
\baselineskip=5truemm
\parskip=3truemm
\input epsf
\def\bye{\vfill\eject}
\def\P{I\!\!P}
\def\half {{\textstyle{1\over 2}}}

\centerline{\Blue{\big The total cross section at the LHC$^{\dag}$\footnote
{}{$^{\dag}${\rm Lectures at School on QCD, Calabria, July 2007}}}}
\bigskip
We do not have the ability to perform precise calculations of
long-range strong interaction effects, because the effective QCD
coupling is not small and so we cannot use perturbation theory.
Nevertheless, I will show that we know a lot, though not nearly
enough. As a measure of our lack of knowledge, the
best prediction for the total cross section at LHC energy is:
$$
\BrickRed{\sigma^{\hbox{{\sevenrm LHC}}}=125 \pm 25~\hbox{mb}}
$$

\bigskip
This set of lectures is about the long-range strong interaction at high energy.
Much of what I know about this subject comes from my long collaboration with
Sandy Donnachie. A few years ago we wrote a book
about it, together with colleagues from Heidelberg University: 
Sandy Donnachie, G\"unter Dosch, Peter Landshoff and Otto Nachtmann,
{\it Pomeron physics and QCD}, Cambridge University Press (2002). 
Most of the material in these lectures
is taken from the book, and 
references
to papers and data may be found in it. 
\bye

\centerline{\Blue{\big Regge theory}}
\vskip 10truemm
Because we do not have the ability to perform precise calculations of
long-range strong interaction effects, 
much of what we know comes from looking at experimental data, and
so these lectures contain rather more data plots than equations. 
In order to use what understanding we have of the theory and apply
it to the data, we have to introduce extra assumptions. I will show
you that, often, making the simplest possible assumptions turns out
to be very successful.

The basic theory is known as Regge theory. It relates together a large
number of different reactions. Among those that I will discuss are
\bigskip
\b Hadron-hadron total cross sections

\b Hadron-hadron elastic scattering

\b Diffraction dissociation

\b Photon and lepton induced reactions
\bigskip

I will show you that we know a lot, but of course not nearly
enough. As a measure of our lack of knowledge, the
best prediction for the total cross section at LHC energy is:
$$
\Blue{\sigma^{\hbox{{\sevenrm LHC}}}=125 \pm 25~\hbox{mb}}
$$
By this I mean that 15 years ago I would have predicted 100~mb with
some confidence, and it is still quite likely that this will prove to
be correct. But a value as large as 150~mb is also quite possible.
\bye

\centerline{\BrickRed{\big History}}
\bigskip
\bigskip
\vbox{\b 1935: Yukawa predicted the existence of the pion --- its\h exchange generates
the static strong interaction}\vskip -17truemm
\line{\hfill\epsfxsize=27truemm\epsfbox[0 0 90 60]{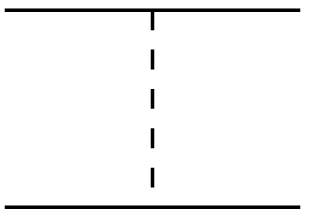}}

\b 1960s: Nearly everybody worked on the applications of Regge theory,
which sums the exchanges of many particles and generates the high-energy
strong interaction

\b The known particles not are enough --- we need to include exchange of another object,
the \underbar{pomeron}

\b 1970s: QCD is discovered --- the BFKL equation generates pomeron exchange as gluon exchange, but it
makes total cross sections rise with energy much faster than is observed 

\b early 1990s: HERA finds a sharp rise of $F_2(x,Q^2)$ at small $x$, apparently described by BFKL pomeron exchange

\b so there are two pomerons: \Blue{soft} (nonperturbative) and \Blue{hard} (perturbative)

\b late 1990s: higher-order perturbative corrections to BFKL exchange spoil
the  calculation --- this problem is not yet solved, and we do not know whether
the hard pomeron is related to the BFKL equation.

\bye

\centerline{\big\Blue{Diffraction}}
\bigskip
The mathematics to study pomeron exchange is sophisticated and\h similar
to that used to study diffraction in optics.

And it leads to elastic scattering differential cross sections
reminiscent of optical diffraction:
\bigskip
\centerline{\epsfxsize=0.4\hsize\epsfbox[50 580 310 770]{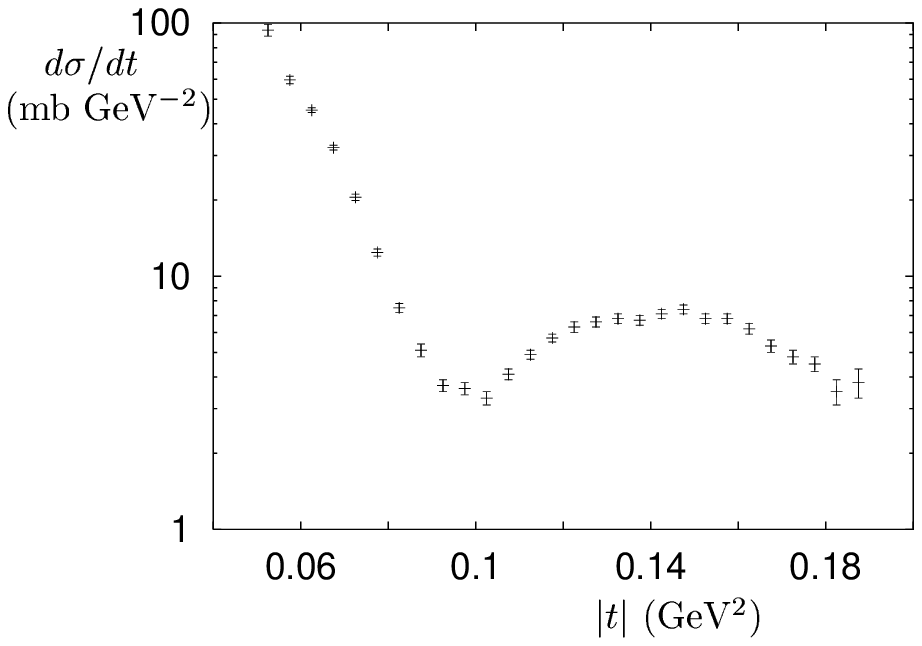}}
\bigskip
\centerline{Elastic $\alpha\alpha$ scattering at 126 GeV CM energy}
\bigskip
\BrickRed{\underbar{But}} "diffractive" processes in particle physics are more complicated than in optics
\bye

\centerline{\big\Blue{Linear particle trajectories}}
\bigskip
Plot of spins of families of particles against their squared masses:
\vskip 5truemm
\centerline{\epsfxsize=140 truemm\epsfbox[63 550 350 755]{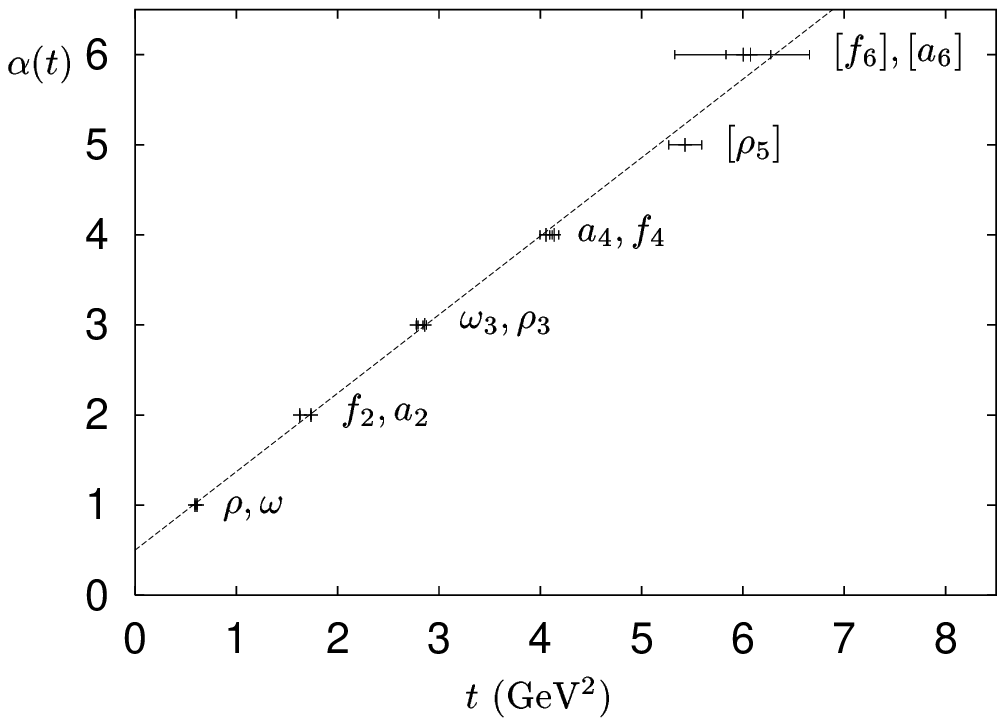}}
\vskip 5truemm
\b 4 degenerate familes of particles:
$
~~\alpha(t)\approx\half +0.9 t
$\h
The particles in square brackets are listed in the data tables, but
there is some uncertainty about whether they exist.

The function $\alpha(t)$ is called a Regge trajectory.
\bye

\centerline{\big\Blue{Regge theory}}
\bigskip
Regge theory sums the exchanges of many particles.
\bigskip
\centerline{\epsfxsize=55 truemm\epsfbox{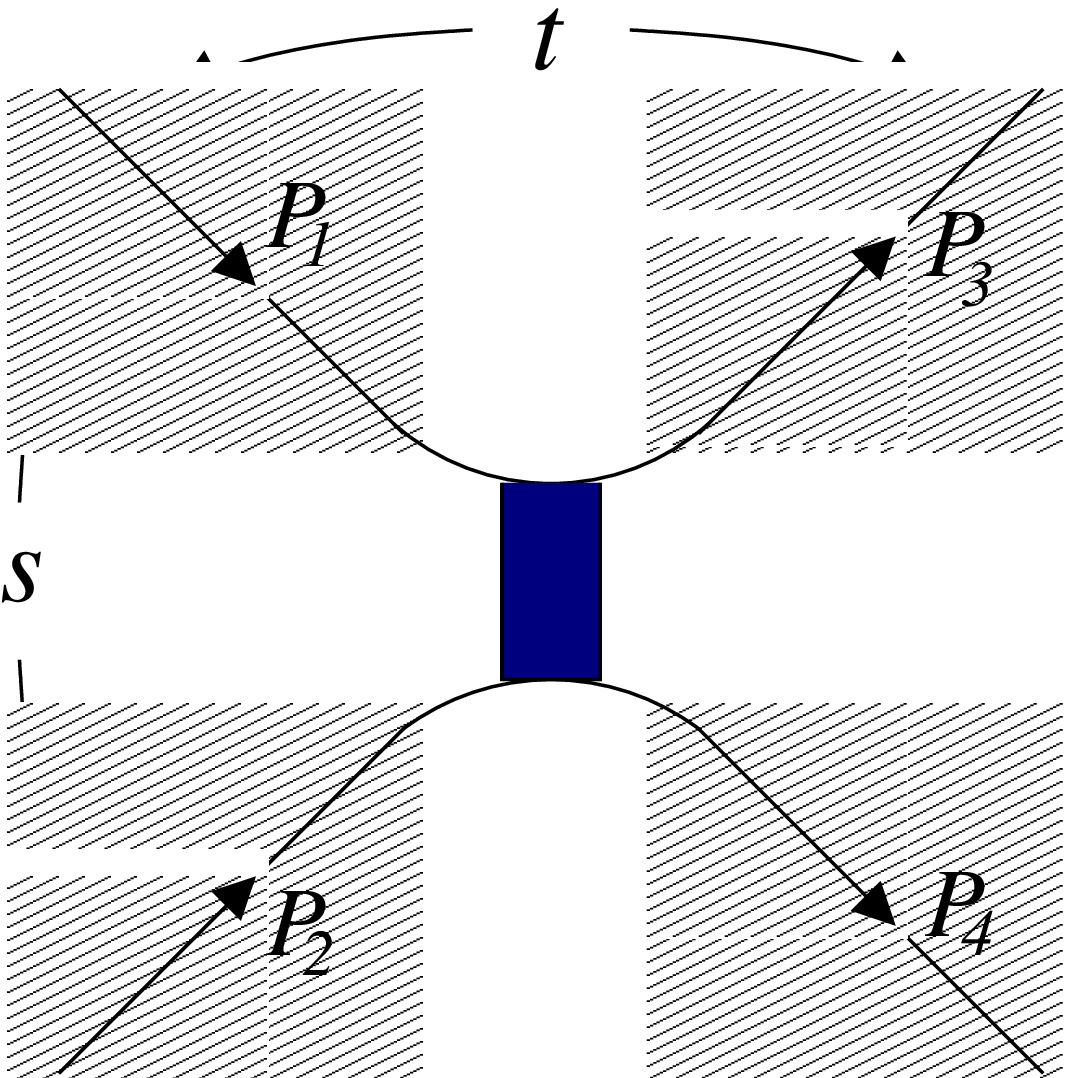}}
\vskip 6truemm
Define\h
$s=(P_1+P_2)^2~$= squared CM energy\h
$t=(P_3-P_1)^2~$= squared momentum transfer

\b At large $s$ but $|t|<<s$ each trajectory $\alpha^{\pm}_i (t)$ contributes
to the amplitude
$$
A^{\pm}(s,t) \sim 
\sum_i\beta^{\pm}_i(t)\Gamma(-\alpha^{\pm}_i (t))
\,\big (1\pm e^{-i\pi\alpha^{\pm}_i(t)}\big )\,(s/s_0)^{\alpha^{\pm}_i(t)-1}
$$
with $\pm$ according to the $C$-parity of the exchange.
We know nothing about the function $\beta^{\pm}_i(t)$, except that it
is \underbar{real}.
$\Gamma(-\alpha^{\pm}_i (t))$ has a pole when $\alpha^{\pm}_i (t)$ is 
a negative integer, but the \underbar{signature factor} $\big (1\pm e^{-i\pi\alpha^{\pm}_i(t)}\big )$ vanishes when $\alpha^{\pm}_i (t)$ is odd/even. 
The signature factor is the only factor that is not real.

The mathematical formalism that is used to derive this high-energy behaviour
of a scattering amplitude was developed by Watson and Sommerfeld in
the middle of the 19th century. It starts with the partial-wave series
for the amplitude, which is a sum over orbital angular momentum values
$\ell = 0,1,2,\dots$. This sum is converted into an integral over
$\ell$, which becomes a continuous complex variable. So the partial-wave
amplitude $a_{\ell}(s)$ becomes a function $a(\ell, s)$ and it has
singularities in the complex-$\ell$ plane. It turns out that a trajectory
$\alpha (t)$ corresponds to a simple pole at $\ell=\alpha (t)$.
As I have indicated, a Regge pole at this point in the complex-$\ell$ plane
contributes a power $s^{\alpha(t)-1}$ to the high-energy behaviour
of the physical amplitude.

Unfortunately, we know that simple poles are not the only singularities
of $a_{\ell}(s)$. It also has branch points, and we do not know enough
about these additional singularities to use them to make well-defined calculations.
It is this difficulty that brought to a halt the intense activity
in Regge theory in the 1960s, and it has still not been solved. I
will discuss it in my last lecture.
\bye

\centerline{\big\Blue{Total cross sections}}
\bigskip
Optical theorem:
$$\sigma^{\hbox{{\sevenrm TOT}}}(s)=\hbox{Im }A(s,t=0)
$$
So each trajectory contributes a fixed power 
$$
s^{\alpha(0)-1}
$$$$
\sim s^{-{1\over 2}} \hbox{ for }\rho,\omega,f_2,a_2 \hbox{ trajectories}
$$

Experiment finds that total cross sections \underbar{rise} gently at large $s$.
So if this is caused by a Regge pole we need another trajectory with $\alpha(0)$ a little $>1$.

\b We call this the \Blue{\underbar{soft pomeron}} trajectory
\bigskip
\b Probably it corresponds to the exchange of \Blue{glueballs}, though
we cannot be sure because the experimental study of the glueball spectrum
is so very difficult.
\bye

\centerline{\big\Blue{Fits to total cross sections}}
\bigskip
\line{\epsfxsize=.5\hsize\epsfbox[60 580 315 770]{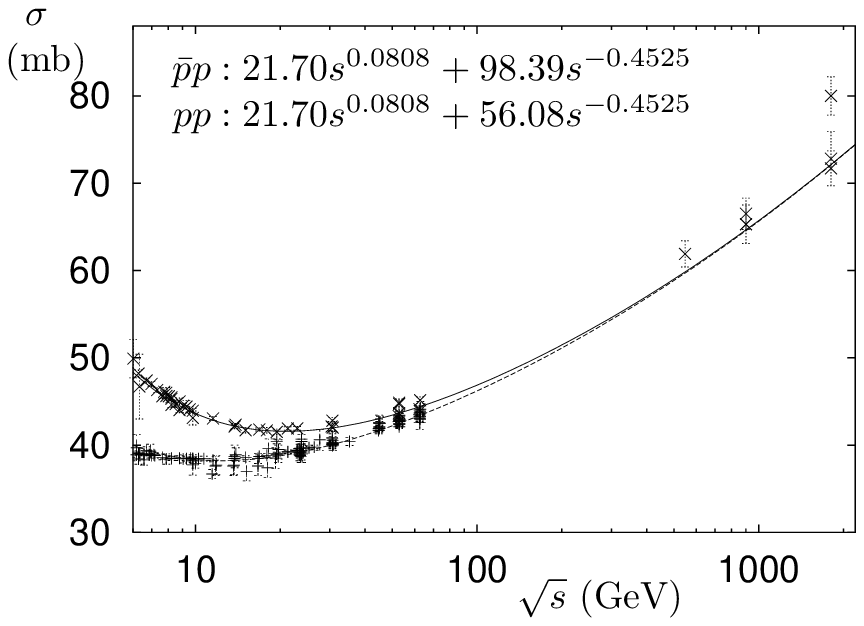}\hfill
\epsfxsize=.5\hsize\epsfbox[60 580 315 770]{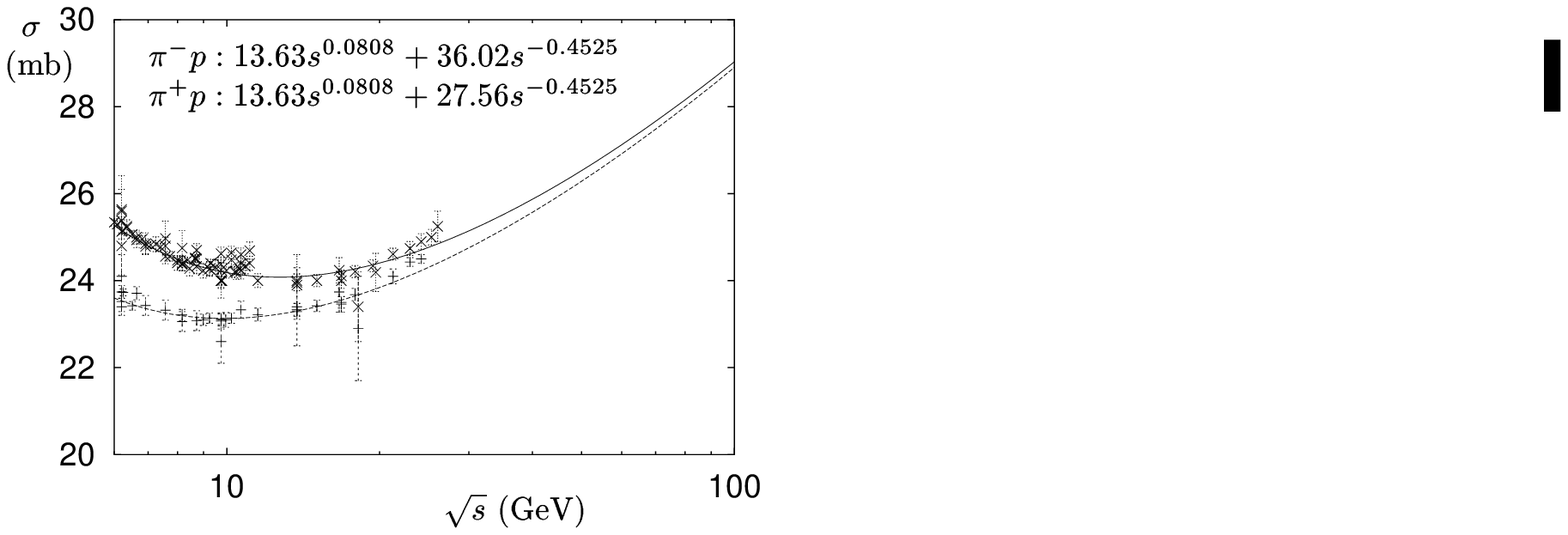}
}
\bigskip
\b We need a pomeron trajectory with $\alpha_{\P}(0)\approx 1.08$. It
couples equally to particles and their antiparticles.
\bigskip
\b Note the significant discrepancy between the two Tevatron measurements
\bigskip
\b Note also that the pomeron's coupling to the pion is about
$2\over 3$ that to the nucleon (\BrickRed{{\sl quark counting rule}})

It is remarkable that such a simple fit works well all the way from
such low energy to very high energies. At the lower-energy end of
the plots, very little can be produced in the final state, just a
very few pions. But as the energy increases, we cross thresholds
for the production of charm, jets and much else. The total cross-section,
however, is completely smooth and seems to be unaware of these
thresholds. 

When Donnachie and I first made these fits to total cross sections,
the higher-energy data from the CERN collider and the Tevatron were
not available, but our predictions based on Regge theory were successful.
\bye

\line{\epsfxsize=.5\hsize\epsfbox[60 580 315 770]{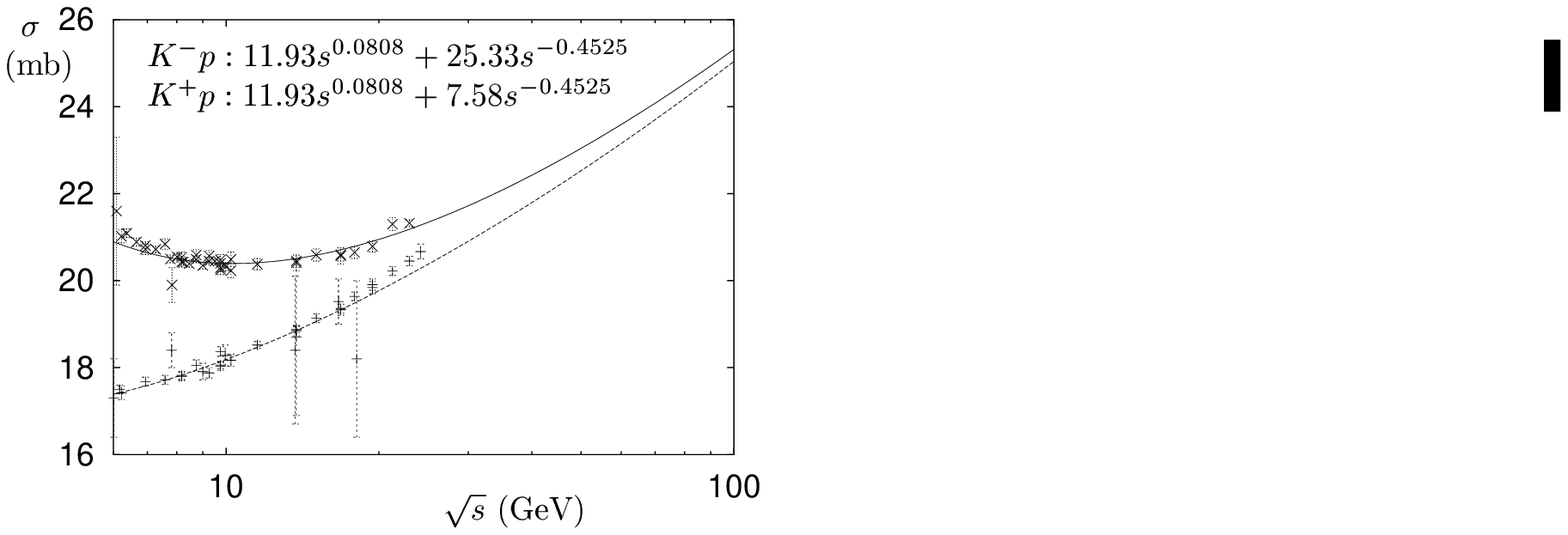}\hfill
\epsfxsize=.5\hsize\epsfbox[60 580 315 770]{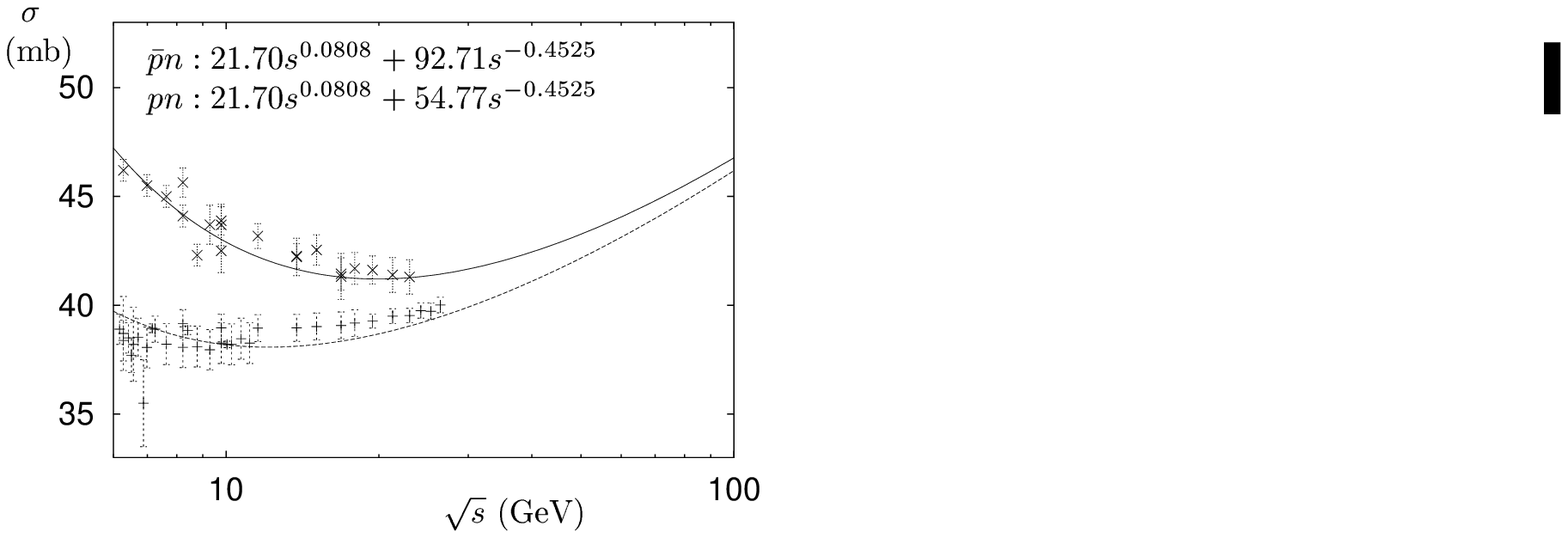}}
\bigskip
The magnitude of the pomeron's contribution to the $Kp$ cross sections
is a little less than to the $\pi p$: the coupling of the pomeron to
a strange quark is about 70\% of that to the light quarks. 
The pomeron's coupling to the $pn$ and $\bar pn$ amplitudes is the same as to 
$pp$ and $\bar pp$: it has the quantum numbers of the vacuum (like the $f_2$).
\vskip 40truemm
\centerline{\epsfxsize=0.5\hsize\epsfbox[60 580 320 690]{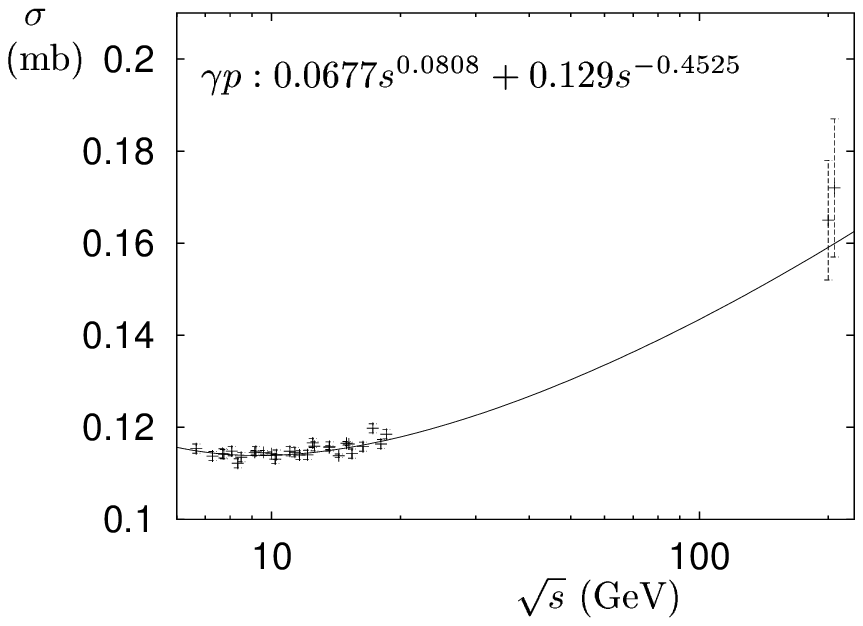}}
\bigskip
Before the HERA measurements of the $\gamma p$ total cross section,
predictions for its value differed very widely. But the prediction
from Regge theory proved to be successful.
\bye

\centerline{\big\Blue{Froissart-Lukaszuk-Martin bound}}

\bigskip
At very large $s$
$$
\sigma^{\hbox{{\sevenrm TOT}}}(s)<{\pi\over m_{\pi}^2}~\log^2(s/s_0)
$$
for some unknown $s_0$ --- probably of the order of 1 GeV$^2$.

At LHC energy, this gives $\sigma^{\hbox{{\sevenrm TOT}}}<$ 4.3 barns

\b \BrickRed{So the bound has little to do with physics!}

Note that the proof depends on the partial-wave unitarity equation 

\epsfxsize=0.6\hsize\epsfbox{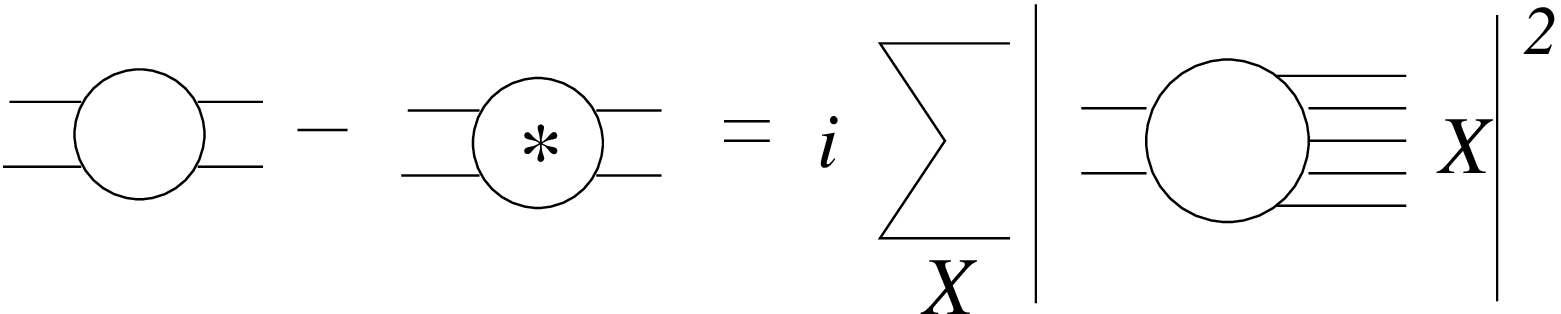}

$$
\hbox{Im }a_\ell (s)=|a_\ell (s)|^2 ~+~\hbox{inelastic terms}
$$
so that
$$
|a_\ell (s)|<1
$$
\b \Blue{{\bf Therefore the bound applies only to hadron-hadron scattering.}}\h

Nevertheless, there is a wide belief that it applies also to photon
and lepton-induced processes. To derive the bound for these processes,
one has to use models or physical intuition. But it is far from certain
that this is reliable under the extreme conditions that will operate
at very high energies. 

\b In principle, the photoproduction cross section
might become very large at high energy, and $F_2(x,Q^2)$ might become very large at small $x$.
\bye

\centerline{\big\Blue{A more stringent constraint}}

Obviously
$$\sigma^{\hbox{{\sevenrm ELASTIC}}}<\sigma^{\hbox{{\sevenrm TOTAL}}}
$$
In fact, unitarity can be used to show that even
$$
\sigma^{\hbox{{\sevenrm ELASTIC}}}<\half ~\sigma^{\hbox{{\sevenrm TOTAL}}}
$$
This is the \Blue{Pumplin Bound}.

\bigskip
Because pomeron exchange alone gives
$$\sigma^{\hbox{{\sevenrm TOTAL}}}\sim s^{\epsilon}~~~~\epsilon\approx 0.08
$$$$
\hbox{and}~~~{d \sigma\over dt}^{\hbox{{\sevenrm ELASTIC}}}\Big |_{t=0}
\sim s^{2\epsilon}~~~~~~~~~~~~~~
$$
it violates the bound at large $s$.
\bigskip
More about this later.
\bye

\centerline{\big\Blue{Elastic scattering}}

Assume the pomeron trajectory is linear (like $\rho,\omega,f_2,a_2$):
$$
\alpha _{\P}(t)=\epsilon _{\P} + \alpha' _{\P}t
$$

\centerline{\epsfxsize=80truemm\epsfbox{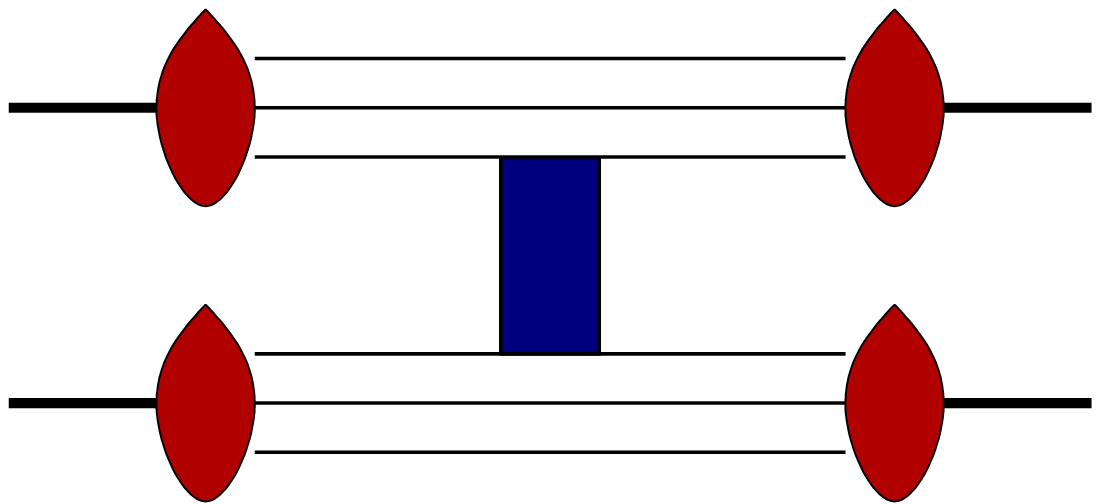}}

\b I pointed out that total-cross-section data suggest the quark counting rule: the pomeron seems to couple to the separate
quarks in a hadron with a $\gamma^{\mu}$ coupling (as expected if
pomeron exchange is two-gluon exchange).
$$
A_{\P}^{qq}(s,t)\sim \bar\beta_{\P}(t)
(\bar u_3\gamma ^{\mu}u_1)(\bar u_4\gamma _{\mu}u_2)
e^{-{1\over 2}i\pi\alpha _{\P}(t)}(\alpha' _{\P}s)^{\alpha _{\P}(t)-1}
$$
\b The two-gluon-exchange model would make $\bar\beta_{\P}(t)$ essentially
constant:\h
\hbox{$
~~\bar\beta_{\P}(t)=\beta _{\P}^2
$}

\b For coupling to nucleons, we need isosinglet (sum of proton and neutron)
$C=+$ Dirac and Pauli form factors
$F_1(t)$ and $F_2(t)$. Assume they are the same as for $C=-$ photon exchange
(this is unlikely to be quite right!). 

\b Then $F_2(t)$ is very small (at $t=0$ it is the sum of the anomalous
magnetic moments of the proton and neutron, $1.79 - 1.9$). This means
pomeron exchange does not flip the nucleon helicity, which is found to be true.

So

$$
{d\sigma\over dt}={{[3 \beta _{\P}F_1(t)]^4\over 4\pi}}  (\alpha' _{\P}s)^{2(\epsilon _{\P} +\alpha' _{\P}t)}
$$

$\beta _{\P}$ and $\epsilon _{\P}$ are known from $\sigma ^{\hbox{{\sevenrm TOT}}}$\h

\b The only free parameter is $\alpha' _{\P}$ 
\bye

Fix $\alpha'$ from the very-low-$t$ data at some energy, say $\sqrt s=$53~Gev:
\bigskip
\centerline{\epsfxsize=0.55\hsize\epsfbox[60 550 360 760]{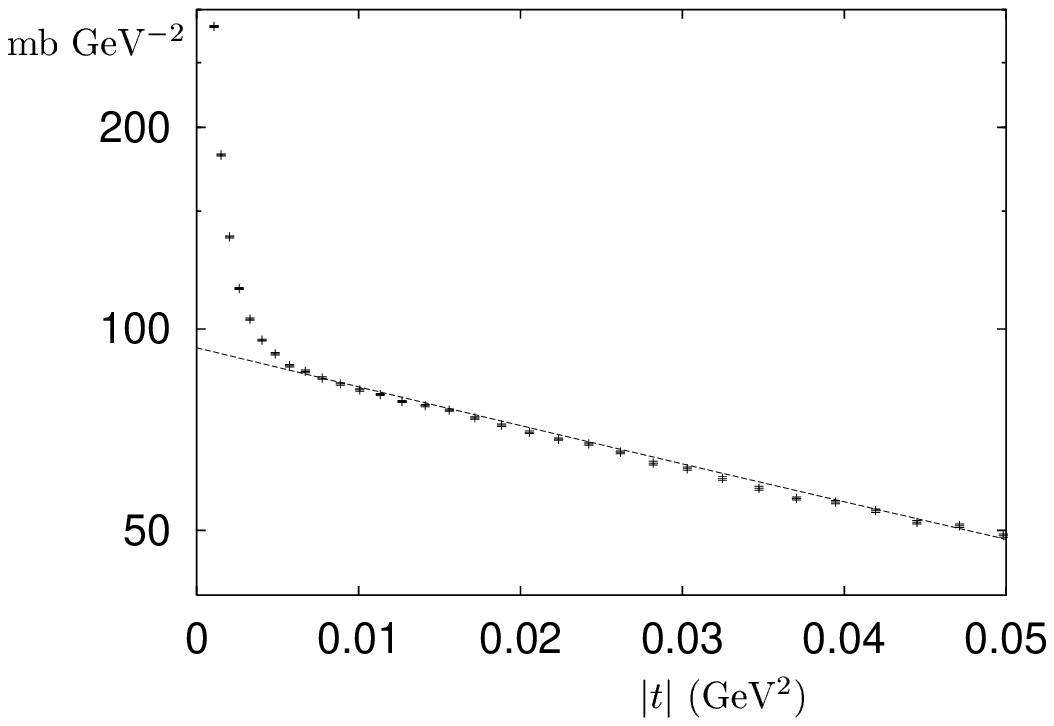}}
\bigskip
\b Determines $\alpha'=0.25$ GeV$^{-2}$

\b Then the formula works well out to larger $t$ at the same energy: 
\bigskip
\centerline{\epsfxsize=0.55\hsize\epsfbox[60 550 360 760]{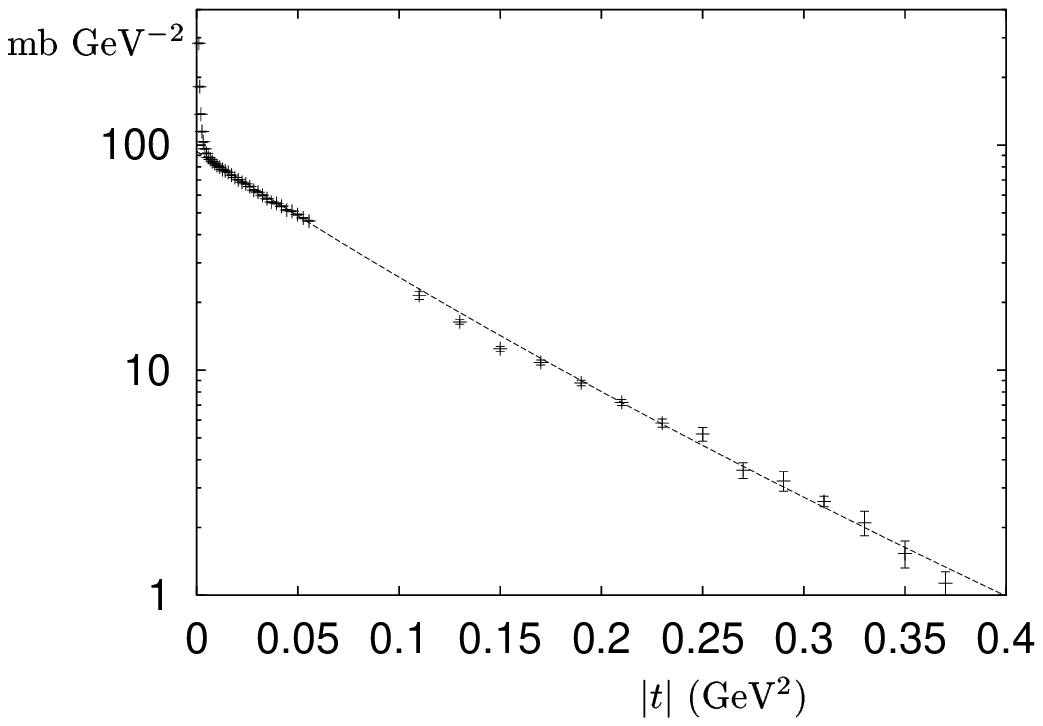}}
\bigskip
It also fits well to $pp$ and $p\bar p$ elastic scattering data
at all other available energies.\h
Because $F_1(t)$ is raised to the power 4 in the formula, this gives
a good test that it is the correct form factor, but why this should be so is
not understood. 

Note that the curves do not include photon exchange, which contributes
significantly at very small $t$.
\bye

\centerline{\big\Blue{Shrinkage of the forward peak}}
\bigskip
Because the formula contains the factor
$\exp(2\alpha' t \log(\alpha' s))$, the contribution of pomeron
exchange to the forward peak in $d\sigma /dt$ becomes steeper
as the energy increases:
\bigskip
\centerline{\epsfxsize=0.6\hsize\epsfbox[50 500 470 770]{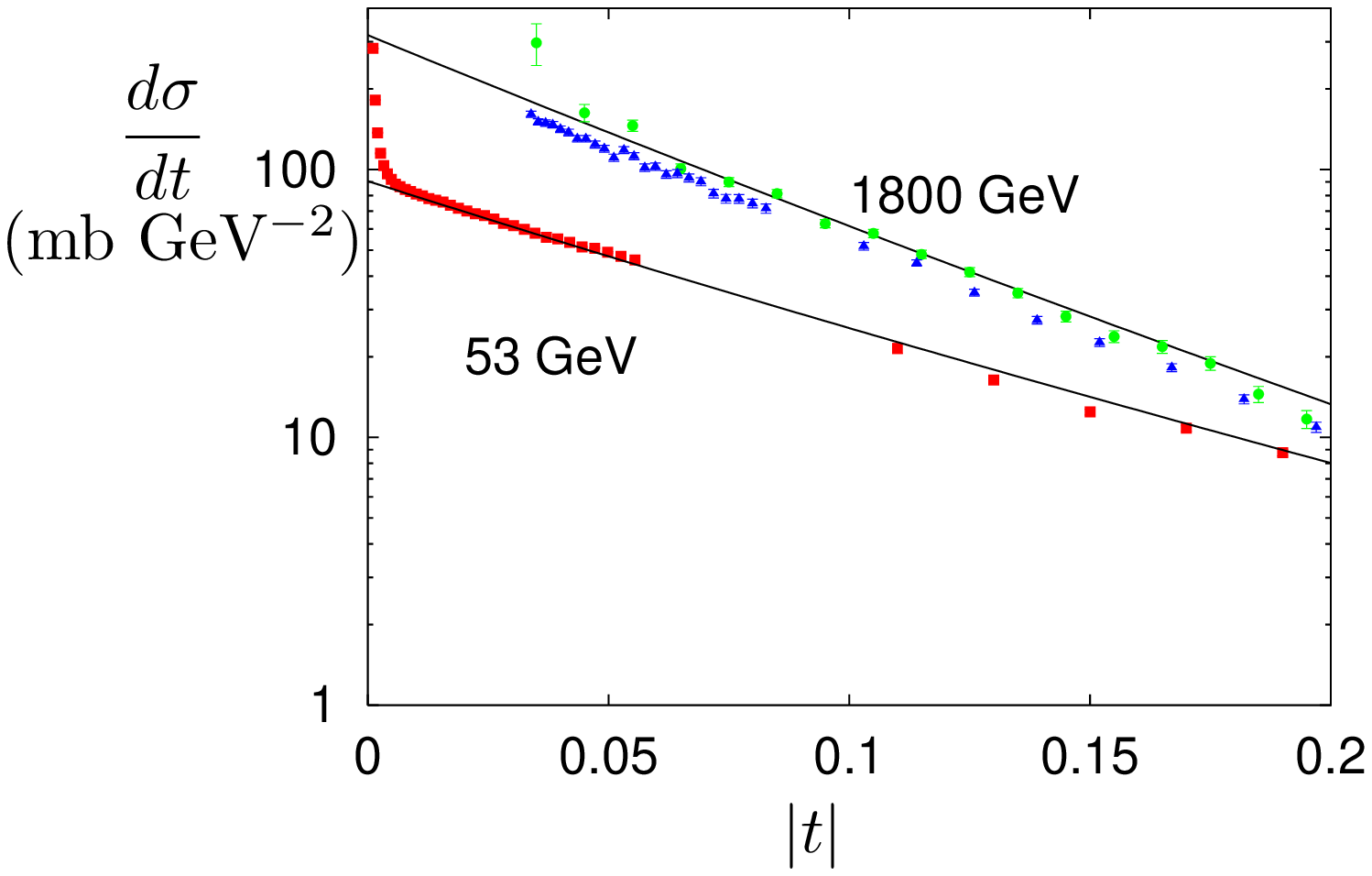}}
\bigskip
Note again the discrepancy between the data from the two Tevatron
experiments.
\bye

\centerline{\big\Blue{$\pi p$ elastic scattering}}

The form factor $F_1(t)$ is raised to the power 4 in the $pp$-%
scattering formula
because the pomeron couples to each of the two protons, so that
$[F_1(t)]^2$ appears in the amplitude, and one has to square the
amplitude to get $d\sigma/dt$. $F_1(t)$ is multiplied by 3
because, according to the quark counting rule, the pomeron couples
to single quarks in the proton and there are three of them. 
If we want to extend the formula to $\pi p$ scattering, we must
replace $[3F_1(t)]^4$  with $[3F_1(t)]^2~[2F_\pi (t)]^2$, where
$F_\pi (t)$ is the elastic form factor of the pion. This has been
measured:
\bigskip
\centerline{\epsfxsize=0.52\hsize\epsfbox[66 604 300 770]{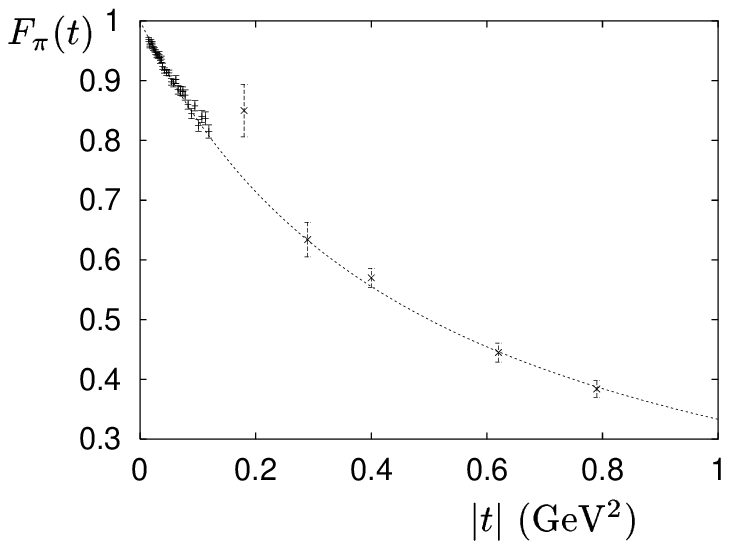}}
\bigskip
and so we obtain for $\pi p$ scattering (at $\sqrt{s} = 19.4$ GeV)
\bigskip
\centerline{\epsfxsize=0.52\hsize\epsfbox[66 604 300 770]{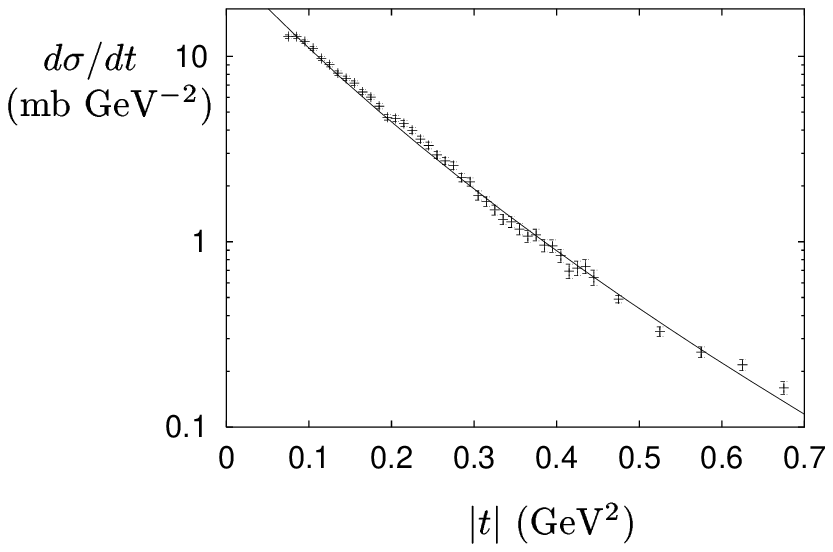}}
\bigskip

The curve has no free parameters!
\bye

\centerline{\big\Blue{Exclusive $\rho$ photoproduction}}
\bigskip
To calculate the amplitude for $\gamma p\to\rho p$,
use vector dominance and assume the $\rho$ behaves like the pion. 
\bigskip

\centerline{\epsfxsize=0.55\hsize\epsfbox[45 564 340 770]{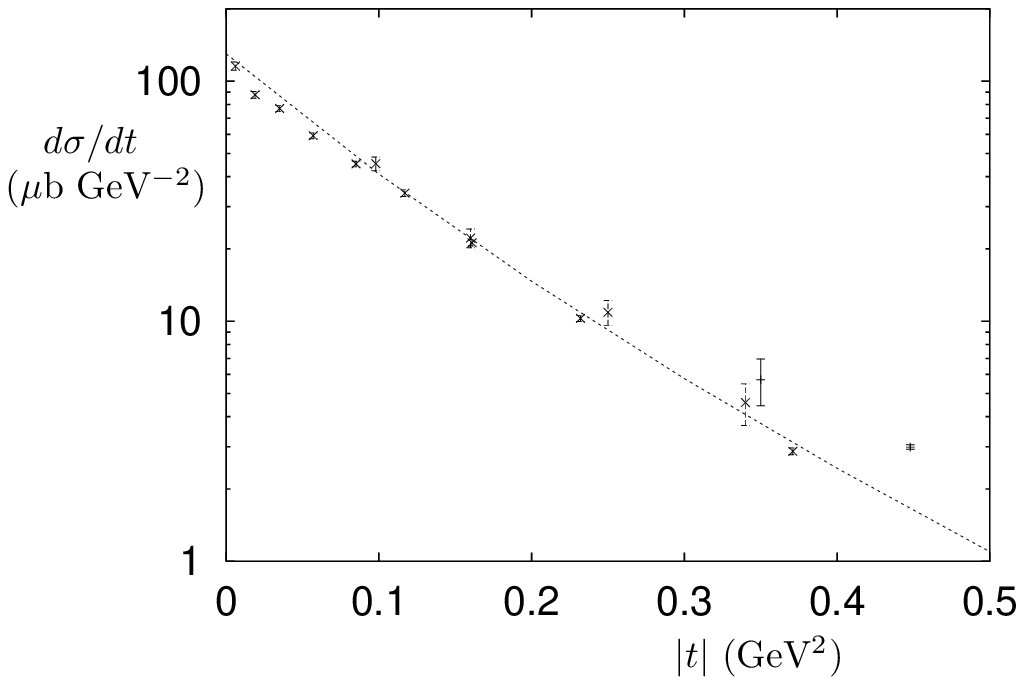}}
\bigskip
This calculation contains no free parameters!
The data are at {$W=$ 71.7 and 94 GeV from ZEUS.

\b Note that both pomeron exchange and vector-meson exchange are
included in this calculation.
\bye

	\centerline{\Blue{{\bigit pp} {\big elastic scattering at large $t$}}}
\bigskip
For $|t|$ greater than about 3 Gev$^2$, the data are consistent
with being energy-independent and fit well to a simple power of
$t$:
$$
d\sigma/dt = 0.09\,t^{-8}
$$
\smallskip
\centerline{\epsfxsize=0.58\hsize\epsfbox[88 460 475 760]{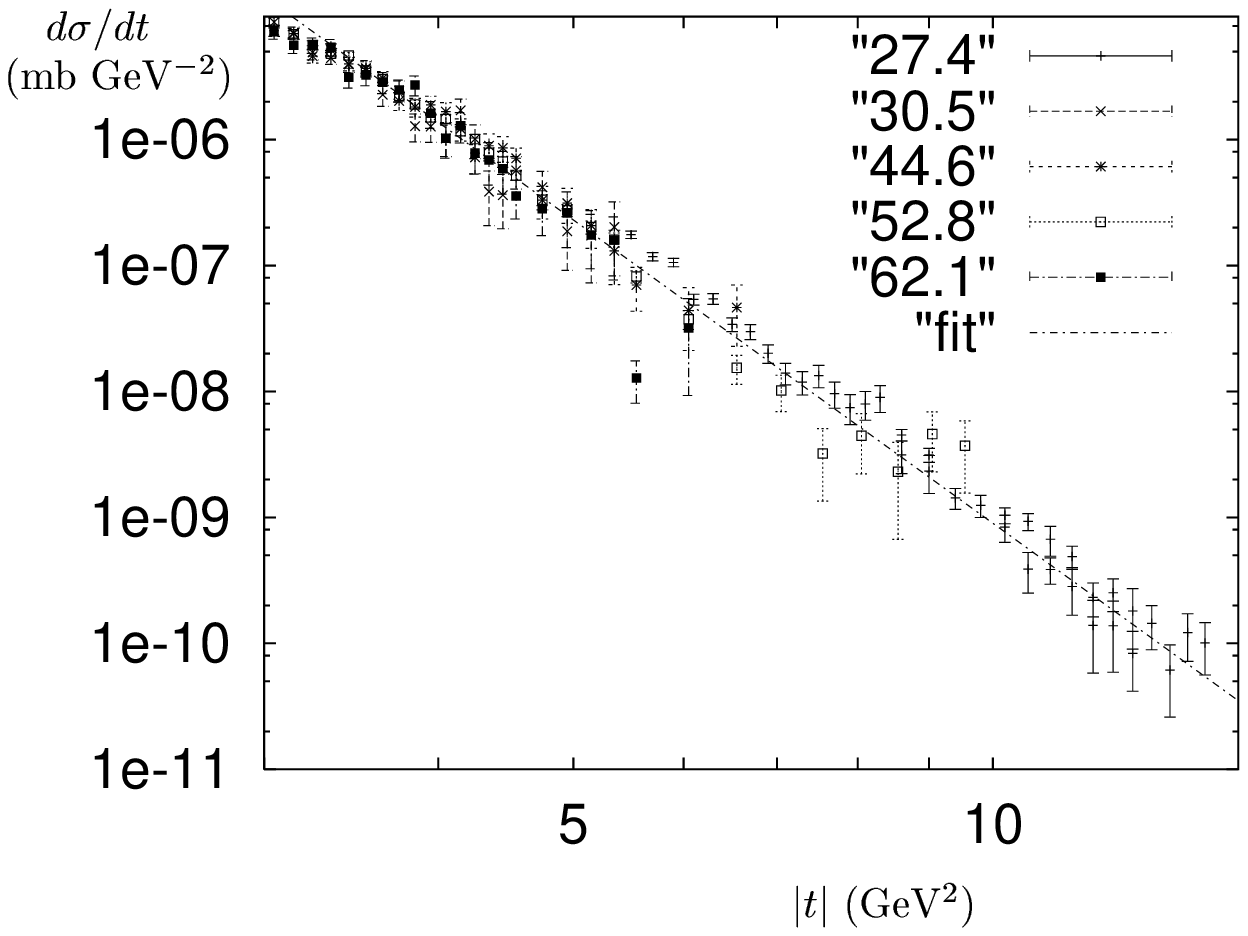}}
\smallskip
This behaviour is what is calculated from triple-gluon exchange:
\bigskip
\centerline{\epsfxsize=0.15\hsize\epsfbox{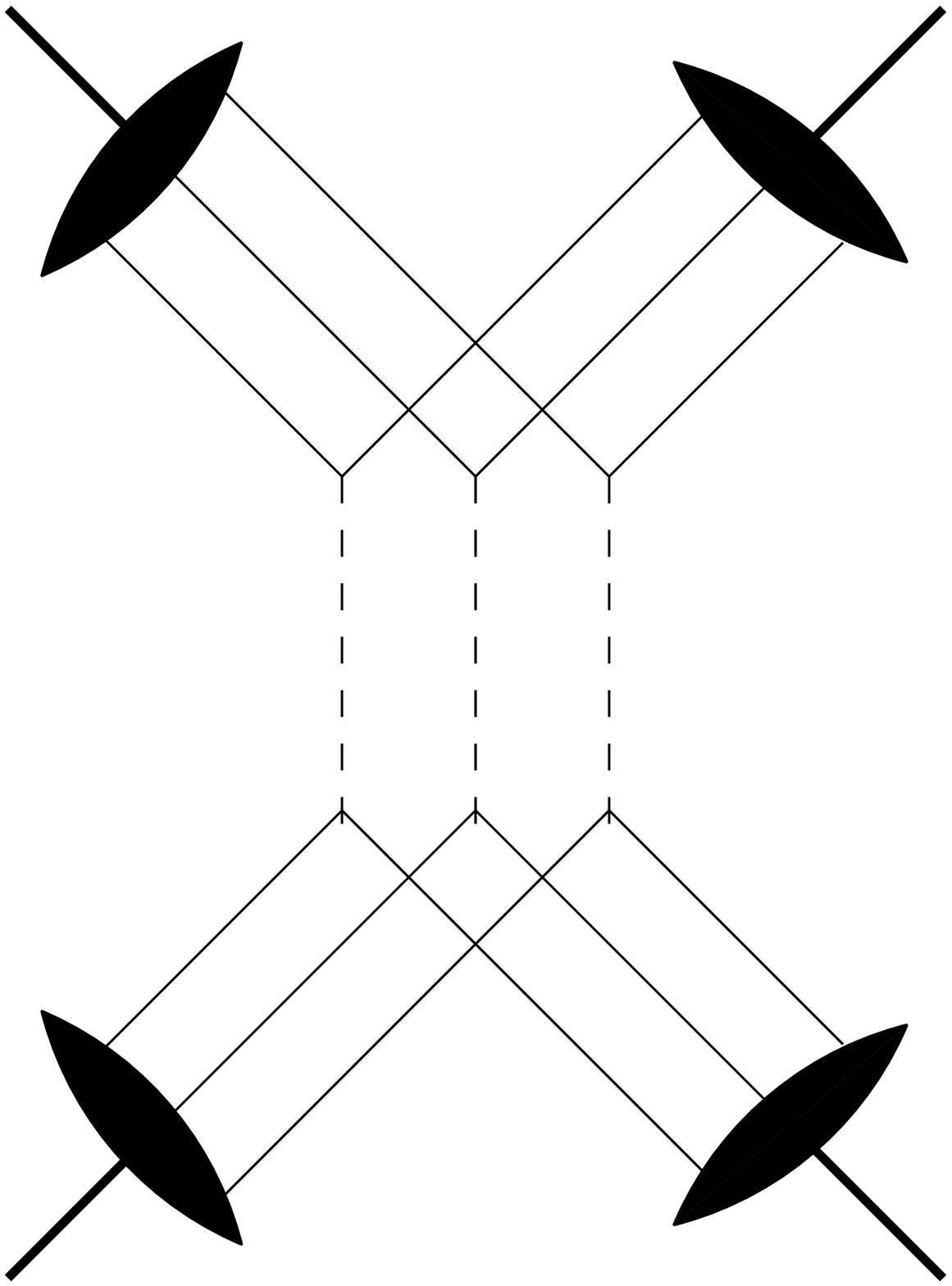}}
\bigskip
It is not understood why this simple mechanism, with no higher-order
perturbative QCD corrections and fixed couplings $\alpha_S$, should
be what is needed. Note, though, that if the proton wave function is
such that on average its momentum is shared equally among the three
quarks, the momentum transfer carried by each gluon is only $t/9$ and
so is quite small. 

As I will explain, the data for $F_2(x,Q^2)$ suggest that there exists
a second pomeron, the hard pomeron, with intercept $\alpha(0)\approx 1.4$.
If we replace the gluons with this, we obtain a contribution that
rises sharply with increasing energy. Although it is not seen in existing data,
it might well become dominant at LHC energy, so that the large-$t$
elastic scattering differential crosss ection might be rather large.

Note that triple-gluon exchange is $C=-1$ --- its contributions to the
$pp$ and $\bar pp$ amplitudes are opposite in sign.
\bye

\centerline{\big\Blue{Smaller values of $t$}}
\bigskip
At smaller values of $t$, the $pp$ elastic scattering data show a
striking dip structure:
\bigskip
\centerline{\epsfxsize=.49\hsize\epsfbox[35 575 320 775]{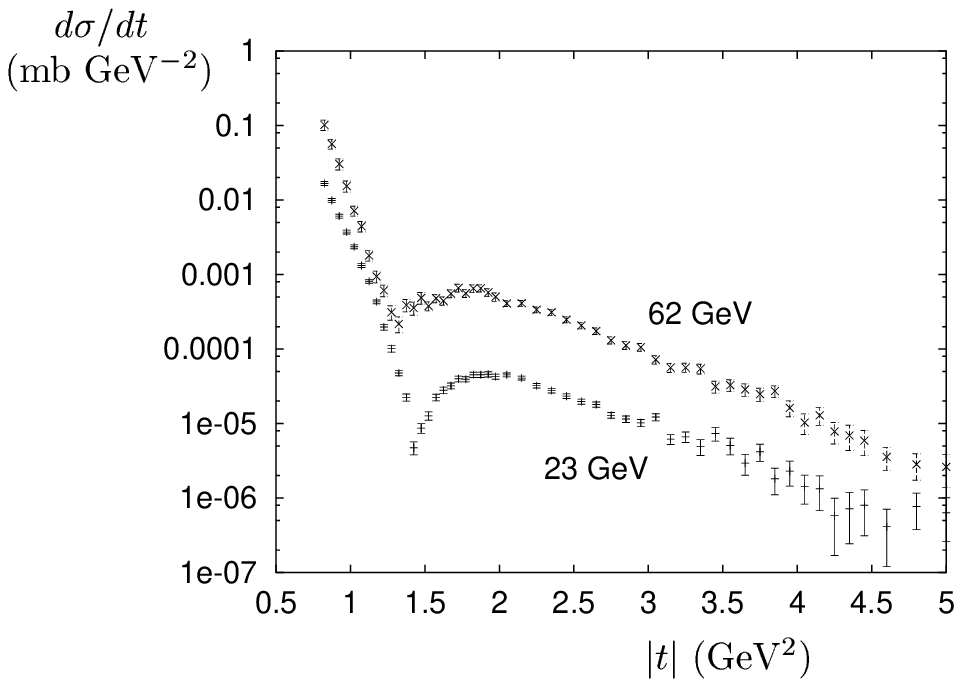}}
\bigskip
(The 62 GeV data are multiplied by 10.)

It is not simple to construct an amplitude that reproduces these dips.
The phase of the contribution from single-pomeron exchange is given by
the signature factor, so that at $-t\approx 1.4$ GeV$^2$ its real
and imaginary parts are of similar magnitude. In order to cancel
both of these at the same value of $t$, additional terms are needed.
One of these is almost certainly the exchange of two pomerons, but its
phase is very different, so at least one other term is needed.
Donnachie and I suggested that this is the triple-gluon-exchange contribution
that is evident at larger values of $t$.

However, for $p\bar p$ scattering this has the opposite sign, which led us
to predict that for this process the dip should be absent. This was
confirmed in measurements at $\sqrt s$=53 GeV:
\bigskip
\epsfxsize=.45\hsize\epsfbox[50 580 310 770]{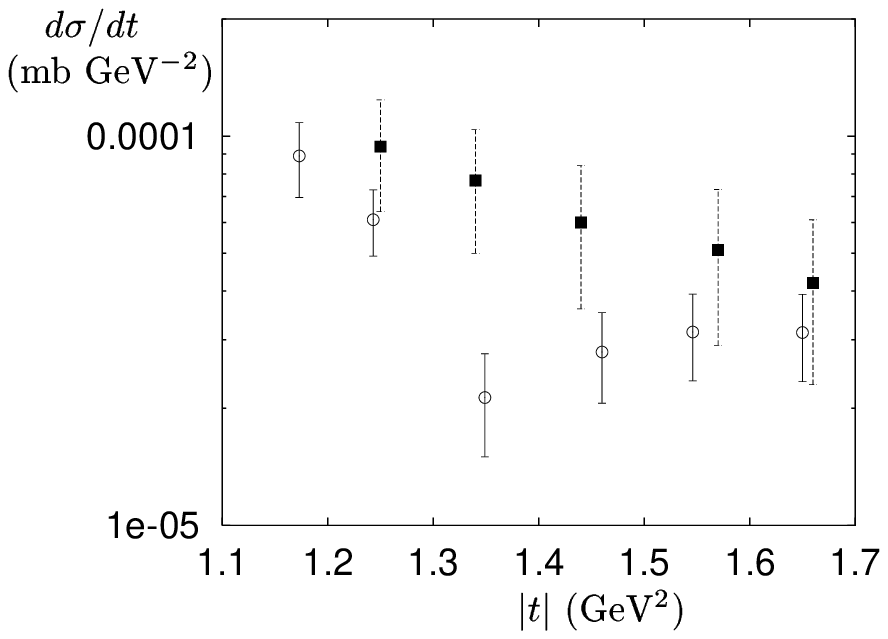}}
\bigskip
\vskip -40truemm
\rightline{\hfill\hfill Upper points $\bar pp$\hfill}\h
\rightline{\hfill\hfill Lower points $pp$\hfill}
\bigskip\bigskip\bigskip
A $C=-1$ exchange term such as triple-gluon exchange is known as an
\underbar{odderon} exchange term.
But there is no sign in the data of odderon exchange at $t=0$.
\bigskip
I will discuss more about the dips later
\bye

\centerline{\big\Blue{Diffraction dissociation}}
\bigskip
This is the name given to the process
$$pp\to pX$$
with the final proton losing only a very small fraction $\xi$ of its 
initial momentum (so therefore there is a large rapidity gap)
\bigskip
\centerline{\epsfxsize=0.5\hsize\epsfbox{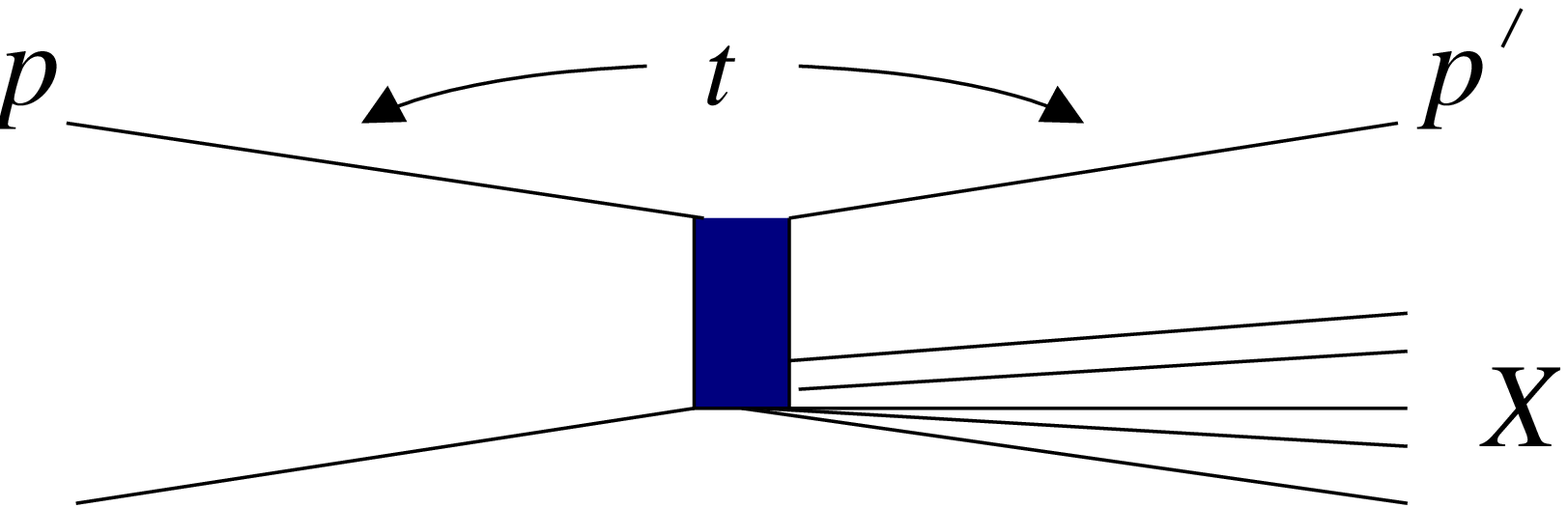}}
\bigskip
Square the amplitude and sum over $X$:
\bigskip
\centerline{\epsfxsize=0.3\hsize\epsfbox{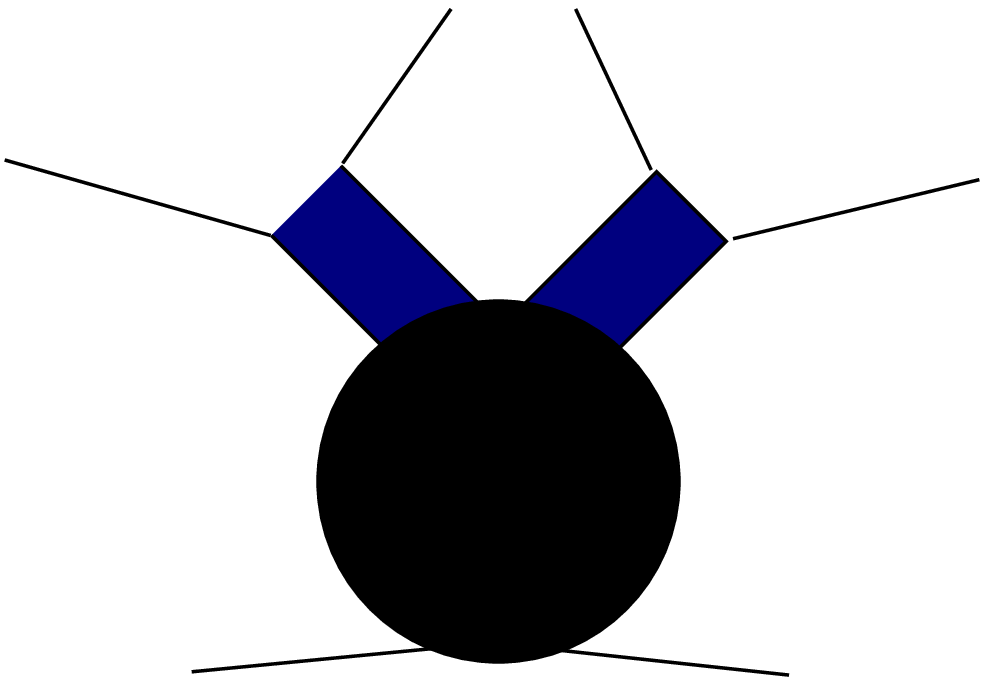}}
\bigskip
We need the imaginary part of the big lower bubble (compare
the optical theorem).
When its energy $M_X$ is large we can apply Regge theory to it and so
 we get the \underbar{triple-regge diagram}
\bigskip
\centerline{\epsfxsize=0.3\hsize\epsfbox{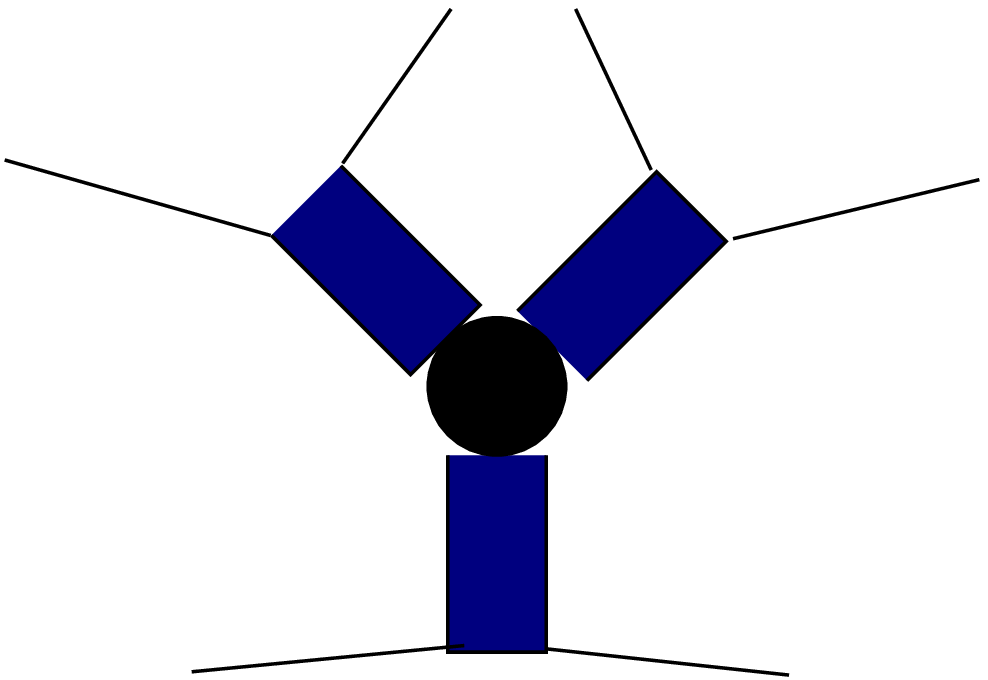}}
\bye

\epsfxsize=0.3\hsize\epsfbox{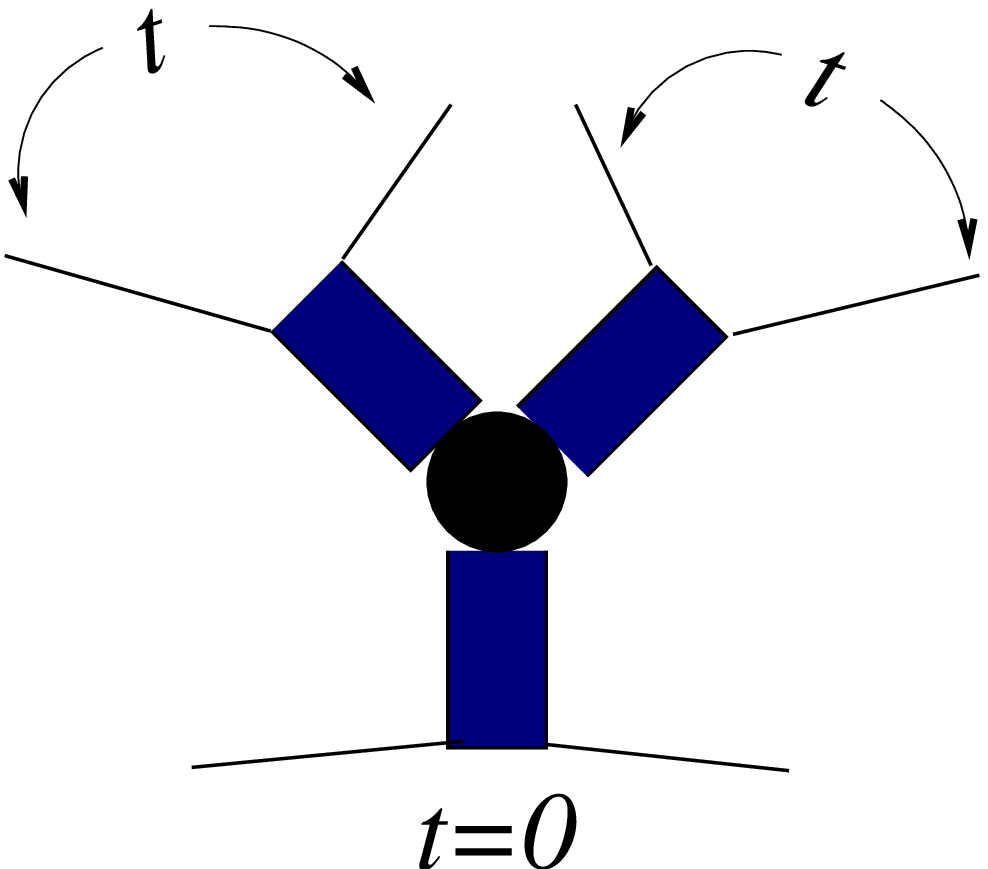}
\bigskip
\b It is not enough just to include the triple pomeron!\h
A very large number of terms need to be considered:
$$
{\P\P\atop\P}~~~~~~{\P\P\atop f_2}
~~~~~~{f_2\P\atop\P}~~~~~~{\P f_2\atop\P}
~~~~~~{f_2\P\atop f_2}~~~~~~
{\omega\P\atop\omega}~~~~~\dots~
$$
A term $({12\atop 3})$ contributes to ${d^2\sigma /dt\, d\xi}$
$$
f^a_1(t)f^a_2(t)f^b_3(0)G^{12}_3(t)\,
e^{i(\phi(\alpha_1(t))-\phi(\alpha_2(t)))}
\xi^{1-\alpha_1(t)-\alpha_2(t)}
\Big ({{M^2}\over{s_0}}\Big )^{\alpha_3(0)-1}
$$
$$
\phi(\alpha(t))=\left\{\matrix{-\half \pi \alpha(t)&C=+1\cr
                        -\half \pi(\alpha(t)-1)&C=-1.\cr}\right .
$$
\bye

\centerline{\big\Blue{Diffdis data}}
\bigskip
Using a combination of terms, it is not difficult to get good
fits to restricted sets of data. For example:
\bigskip
\line{\epsfxsize=0.4\hsize\epsfbox{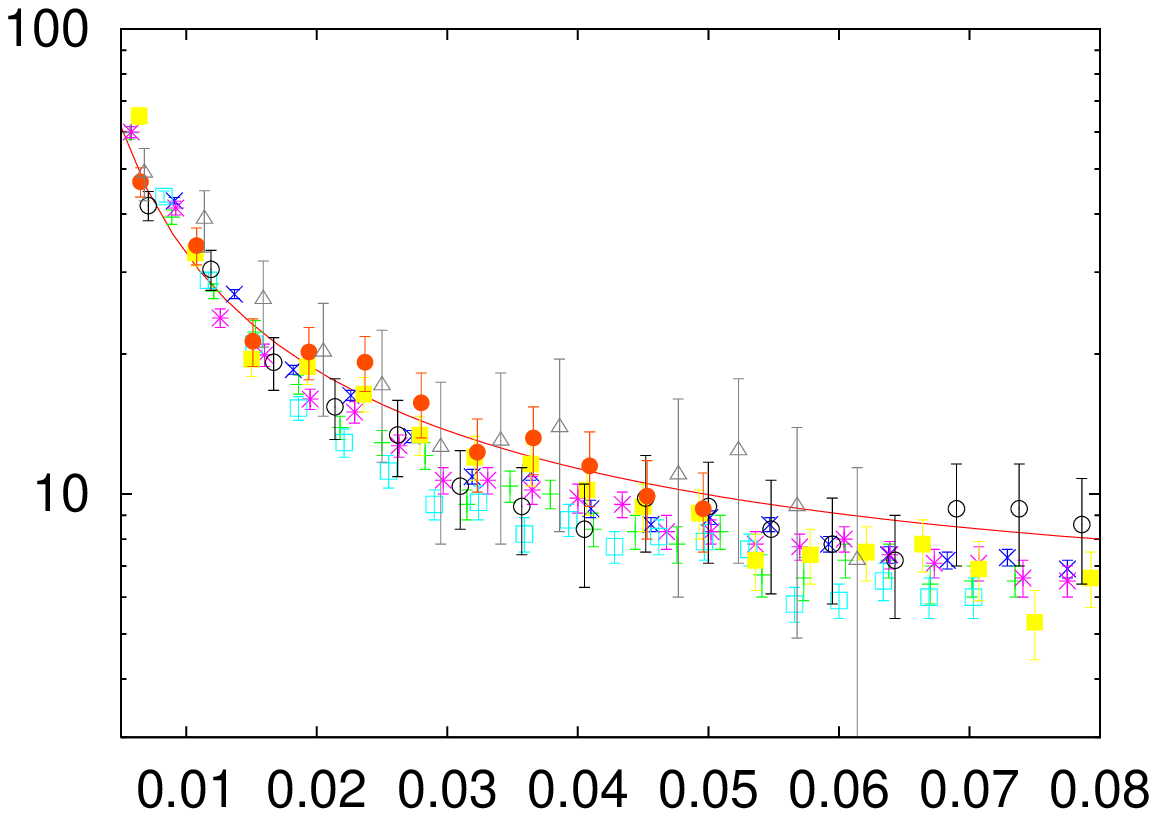}
\hfill
\vbox{\vskip -80 truemm Data at $-t=0.25$ GeV$2$ plotted against $\xi$\h from two experiments at the CERN~ISR\h at $\sqrt s$ from 23 to 38 GeV,\h with a simple triple-regge fit}}
\bigskip
However, there are other data which are much more difficult to accommodate,
for example these fixed-target data:
\bigskip
\centerline{\epsfxsize=0.6\hsize\epsfbox[185 590 435 775]{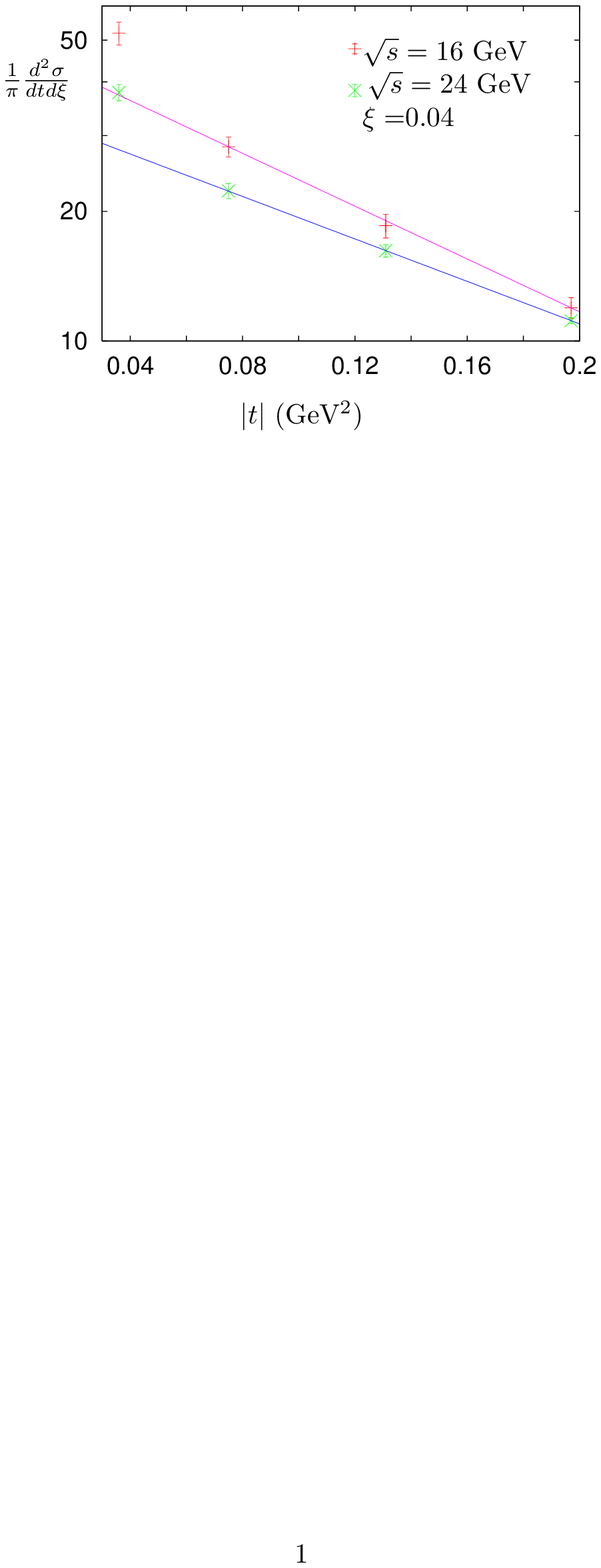}}
\bigskip
Notice two characteristics of these data: the cross section decreases
as the energy increases, and while a simple exponential in $t$
fits well to the three highest-$t$ values at each energy, it fails
to describe the lowest-$t$ points.

\b All too often, fits are made to a restricted set of data using
only pomeron exchange, with the result that conclusions are reached
that are almost certainly wrong.
\bye

The data at higher energies present severe problems. Sadly, the
UA4 data for $d^2\sigma/dt~d\xi$ have been lost.
CDF publish no data points, just their fit.

UA4 data for $d\sigma/dt$ survive,
that is $d^2\sigma/dt~d\xi$
integrated over a certain range of $\xi$. The result of integrating
the CDF fit similarly is as follows:
\bigskip
\centerline{\epsfxsize=0.6\hsize\epsfbox[160 520 450 750]{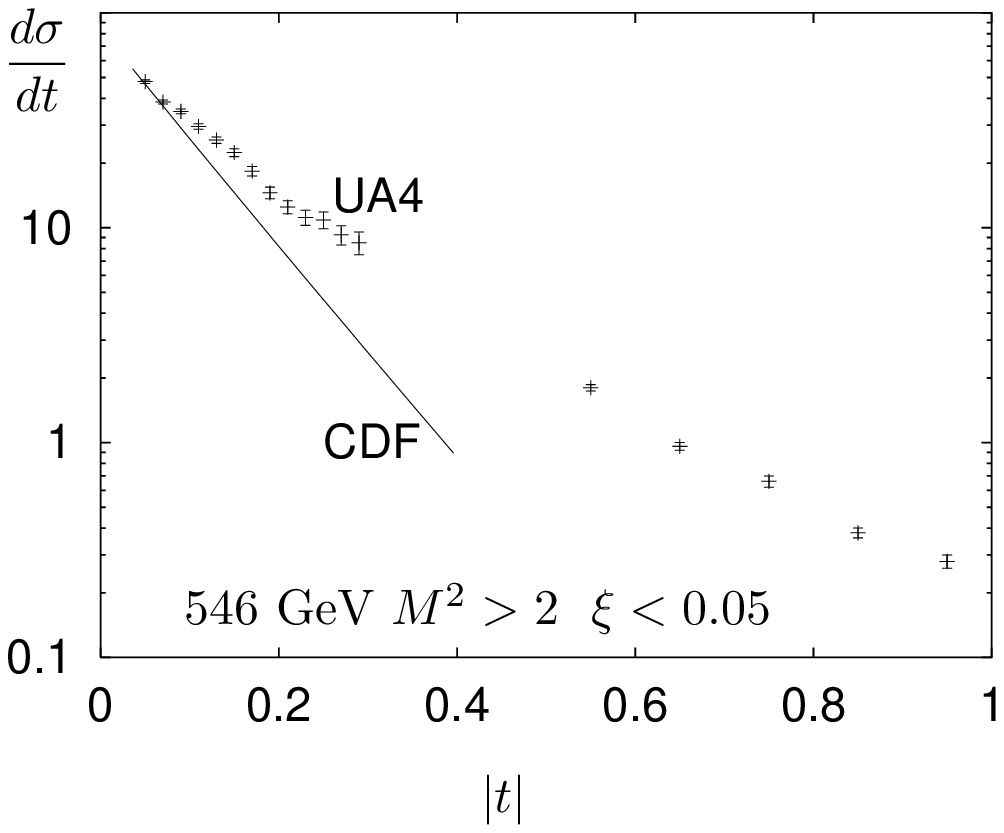}}
\bigskip
\Blue{There is an obvious problem!}

\bye

\centerline{\big\Blue{Deep inelastic lepton scattering}}
\bigskip
\centerline{\epsfxsize=0.4\hsize\epsfbox{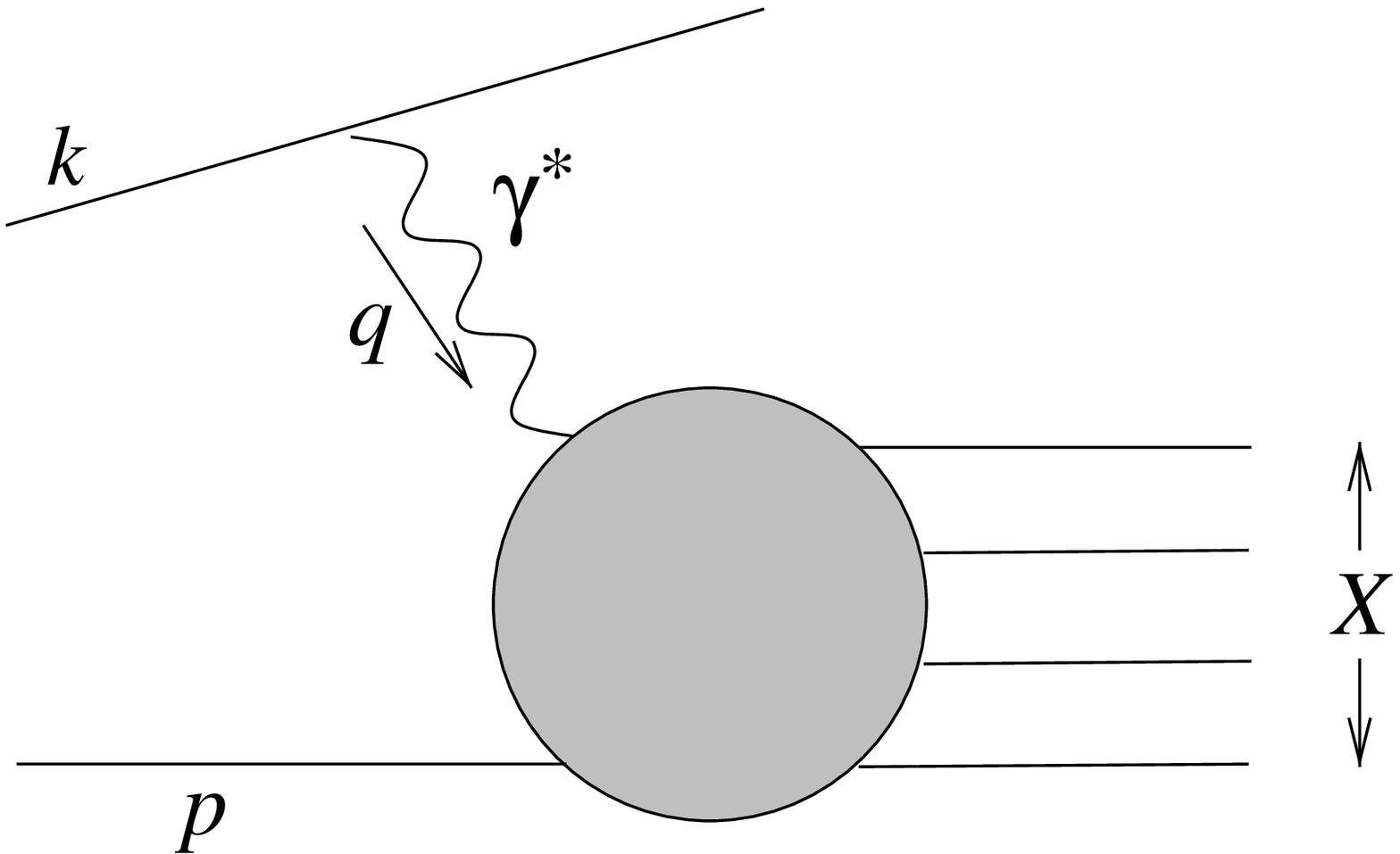}}
\bigskip
Define
$$W^2=(p+q)^2~~~~~\nu=p.q~~~~Q^2=-q^2~~~~x=Q^2/(2\nu)
$$
Soft-pomeron exchange contributes a behaviour $(W^2)^{0.08}$ at fixed $Q^2$.
I have shown how this describes the data well for $Q^2=0$.
But the $W$ dependence gets steeper as $Q^2$ increses: 
\bigskip
\centerline{\epsfxsize=0.6\hsize\epsfbox[90 580 330 775]{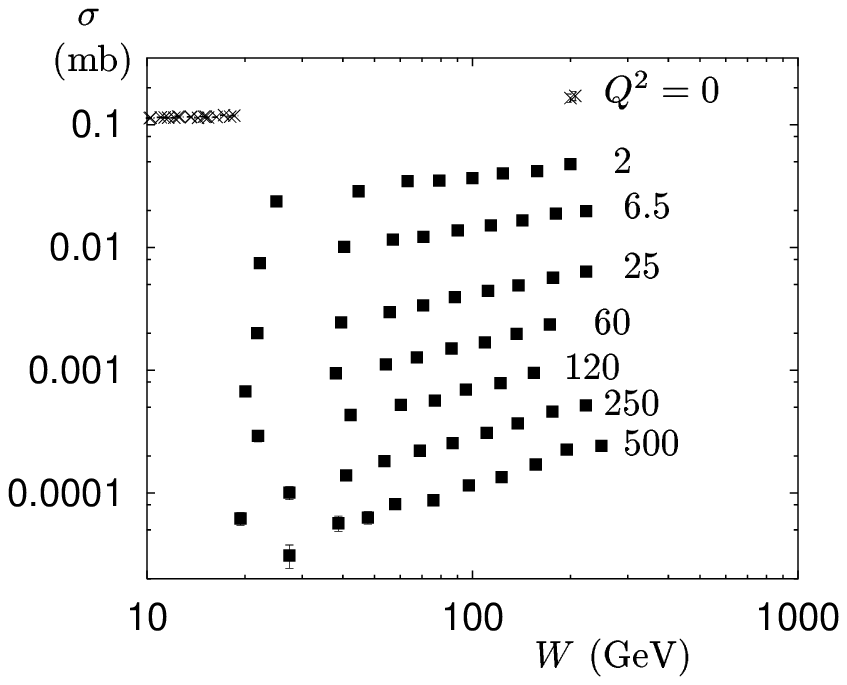}}
\bigskip
\b At large $Q^2$ the behaviour is about $(W^2)^{0.4}$.
\bye

\centerline{\Blue{{\big The theory is not well understood}}}
\bigskip
Define
$$
\sigma(W^2,Q^2)={4\pi^2\alpha_{{\sevenrm EM}}^2\over Q^2}F_2(x,Q^2)
$$

\Blue{Two alternative theoretical possibilities:}
\medskip
\b A $(W^2)^{0.4}$ term is there at all $Q^2$, but at $Q^2=0$ its contribution is very small

\b It is not there at $Q^2=0$, but as $Q^2$ increases it is gradually
generated through perturbative QCD evolution.
\bigskip

The second approach is conventional, but it has a mathematical problem,
as I will explain.
\bigskip
The first leads to the possibility that $\sigma (pp)$ also has an
$s^{0.4}$ term, so that the LHC total cross-section is big.
\bye

\centerline{{\Blue{{\big Simple Regge fit}}}}
\bigskip
Soft pomeron exchange contributes to $F_2(x,Q^2)$ at large $W$ a term that behaves as $(W^2)^{\epsilon_1}$,
with\hbox{$\epsilon_1\approx 0.08$,}. 
This
is equivalent to $(1/x)^{\epsilon_1}$ at large $1/x$, say $1/x > 10^3$.
At smaller values of $1/x$, we must add in $f_2$ and $a_2$ exchange,
and include a multiplicative factor in each term so as to make it
go to 0 appropriately as $x\to 1$. So to begin with let me restrict
the discussion to $x<10^{-3}$.

We have seen that soft-pomeron exchange is not enough, so Donnachie
and I added in another term which we call hard-pomeron exchange. It
behaves as $(1/x)^{\epsilon_0}$, and I have shown that we need
$\epsilon_0\approx 0.4$. In order to pin down the contribution from
this term, we went through a number of steps:

(1) Take
$$
F_2(x,Q^2)=f_0(Q^2)x^{-\epsilon_0}+f_1(Q^2)x^{-\epsilon_1}~~~x<10^{-3}
$$
In order to get information about the unknown functions $f_0(Q^2)$
and $f_1(Q^2)$, choose a value for $\epsilon_0$ somewhat less than 0.4 and
another somewhat greater. Then 
fit the available data at each $Q^2$ at each $Q^2$. 
This gives these outputs for the two functions:
\bigskip
\line{\epsfxsize=0.45\hsize\epsfbox[70 590 300 770]{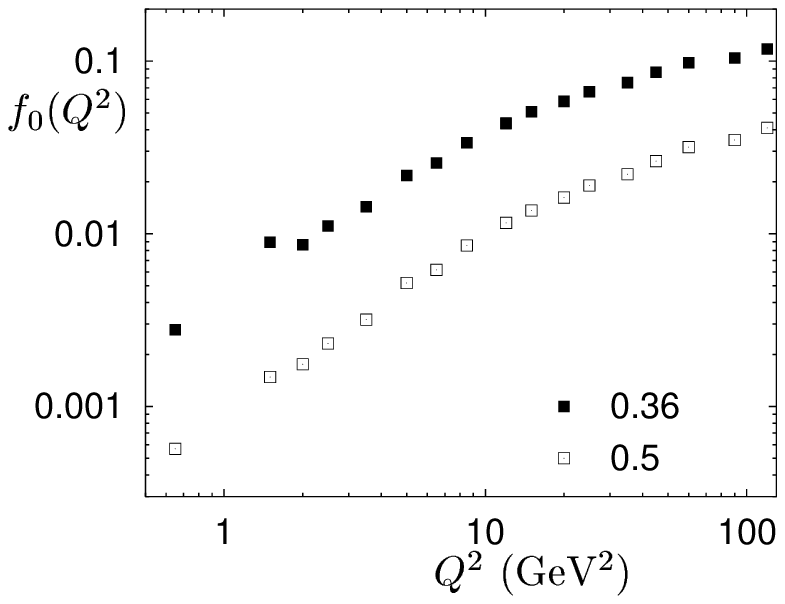}\hfill
\epsfxsize=0.45\hsize\epsfbox[70 590 300 770]{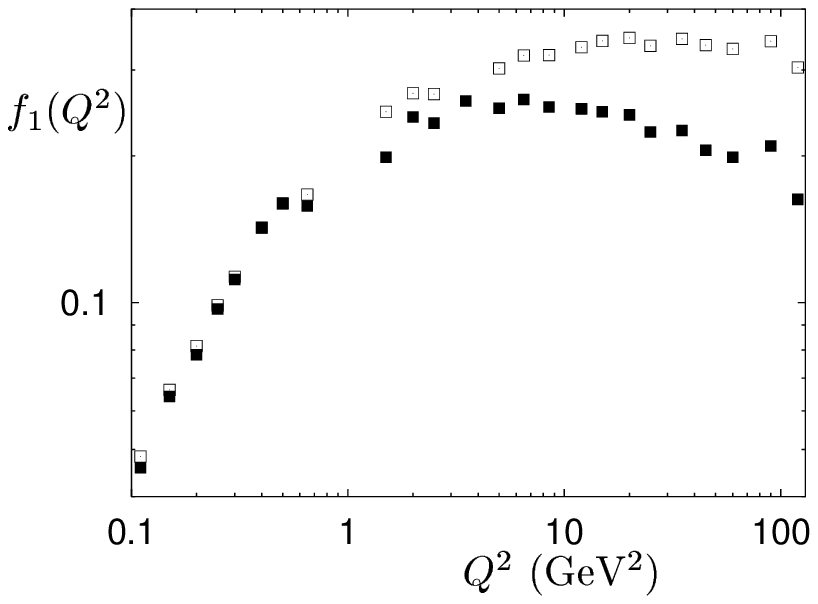}}
\bigskip
The black points are for $\epsilon_0=0.36$, and the  white points $\epsilon_0=0.5$.

In each case, $f_0(Q^2)$ rises steadily with $Q^2$, while $f_1(Q^2)$
either goes to a constant or rises to a peak and then slowly decreases. 
\vfill\eject
(2) This suggests parametrisations of $f_0(Q^2)$ and $f_1(Q^2)$.
Current conservation implies that that near $Q^2=0$ at fixed $W$, $ F_2(x,Q^2)$ vanishes
like $Q^2$.  Therefore
$f_i(Q^2)\sim (Q^2)^{1+\epsilon_i}.~~~$ Take
$$
f_0(Q^2)=A_0\Big ({Q^2\over 1+Q^2/Q_0^2}\Big )^{1+\epsilon_0}
(1+Q^2/Q_0^2)^{\epsilon_0/2}
$$$$
f_1(Q^2)=A_1\Big ({Q^2\over 1+Q^2/Q_1^2}\Big )^{1+\epsilon_1}
$$
For simplicity, this choice makes $f_1(Q^2)$ go to a constant at
large $Q^2$.

Although its contribution for $x<0.001$ is fairly small, we include also
an $f_2,a_2$ exchange term, that is we add
$$
f_R(Q^2)x^{-\epsilon_R}~~~~~~~~~\epsilon_R=-0.4525
$$
and use a similar parametrisation for $f_2(Q^2)$ to that for $f_1(Q^2)$:
$$
f_R(Q^2)=A_R\Big ({Q^2\over 1+Q^2/Q_R^2}\Big )^{1+\epsilon_R}
$$
(3) Now go back and again fit the data for $x<0.001$, this time with
$\epsilon_0$ as one of the free parameters. Include also the photoproduction
data and restrict to $W>6$ GeV.
This gives
$$
\epsilon_0=0.41~~~ Q_0=2.9\hbox{ GeV}~~~~~~Q_1=770\hbox{ MeV}~~~~~~Q_R=465\hbox{ MeV}
$$$$
A_0=0.0022~~~A_1=0.60~~~A_R=1.2
$$
with $ \chi^2=0.95$ per data point (190 data points).

(The values of $A_1,A_2$ are determined largely by the real-photon data.)
"26.tex" 35L, 1236C                                           1,1           Top

\bye

\line{\epsfxsize=0.38\hsize\epsfbox[90 430 290 760]{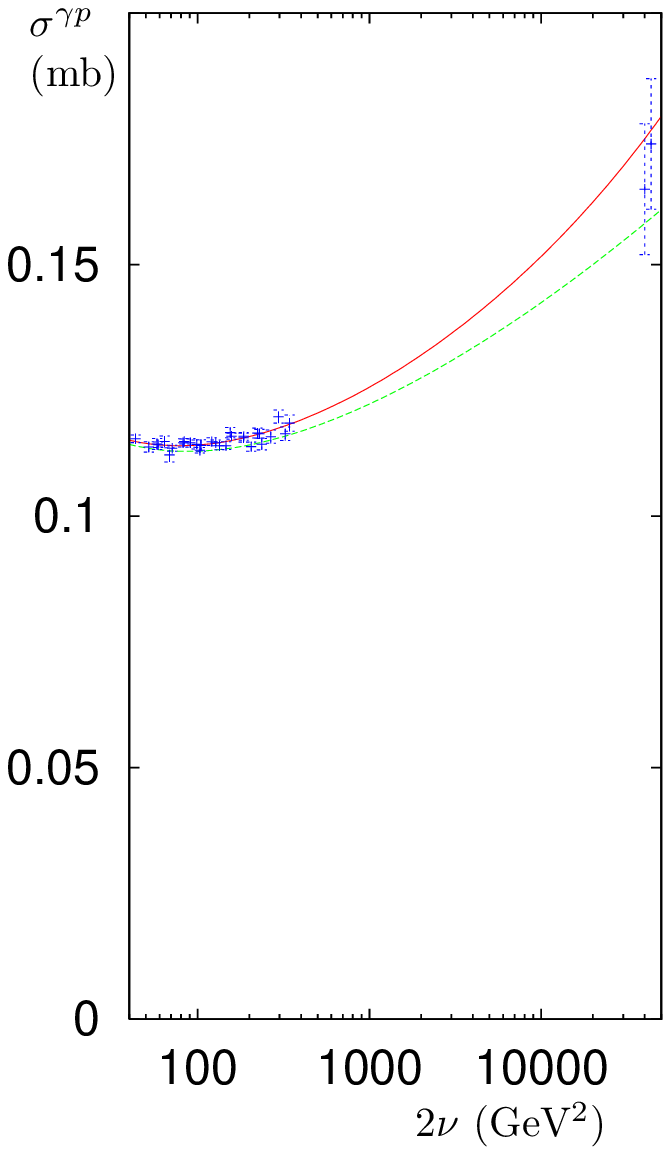}\hfill
\epsfysize=0.55\vsize\epsfbox[65 295 377 755]{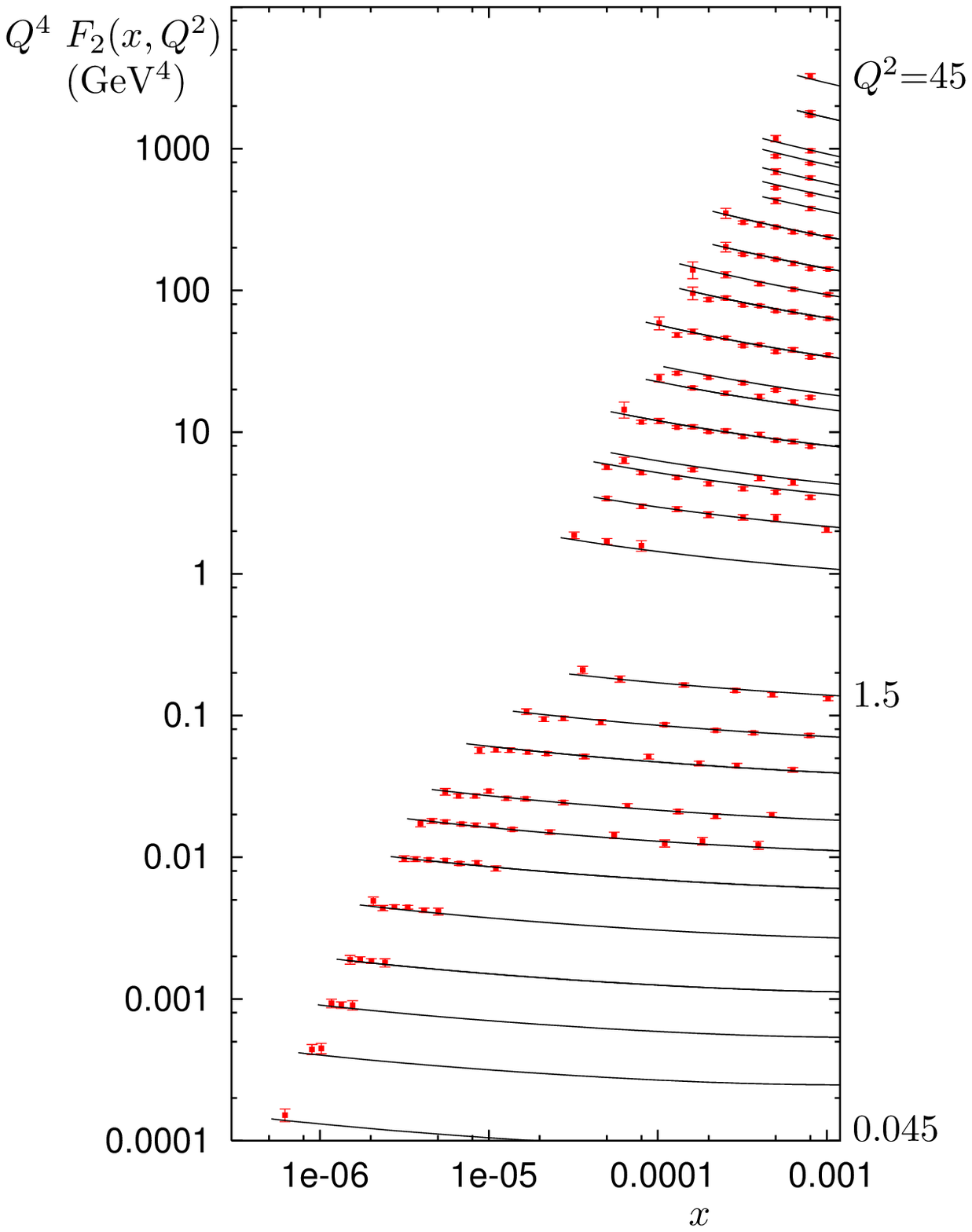}$\phantom{XXXXXXXX}$}
\bigskip
Parametrisations in which $f_0(Q^2)$ falls slowly at large $Q^2$, for example
like $1/Q$, also describe the data well. If only for this reason, the error on
the determination of $\epsilon_0$ is quite large, say
$\epsilon_0=0.41\pm 0.06$.
\bye

(4) There are data at larger $Q^2$ but at larger $x$. Introduce
powers of $(1-x)$ in each term given by the dimensional counting rule.
This is not correct, but better than nothing, and using the values of the
parameters that have already been determined by the data for $x<0.001$
they give surprisingly good fits out to very large $Q^2$:
\bigskip
\centerline{\epsfysize 0.8\vsize\epsfbox[100 300 450 770]{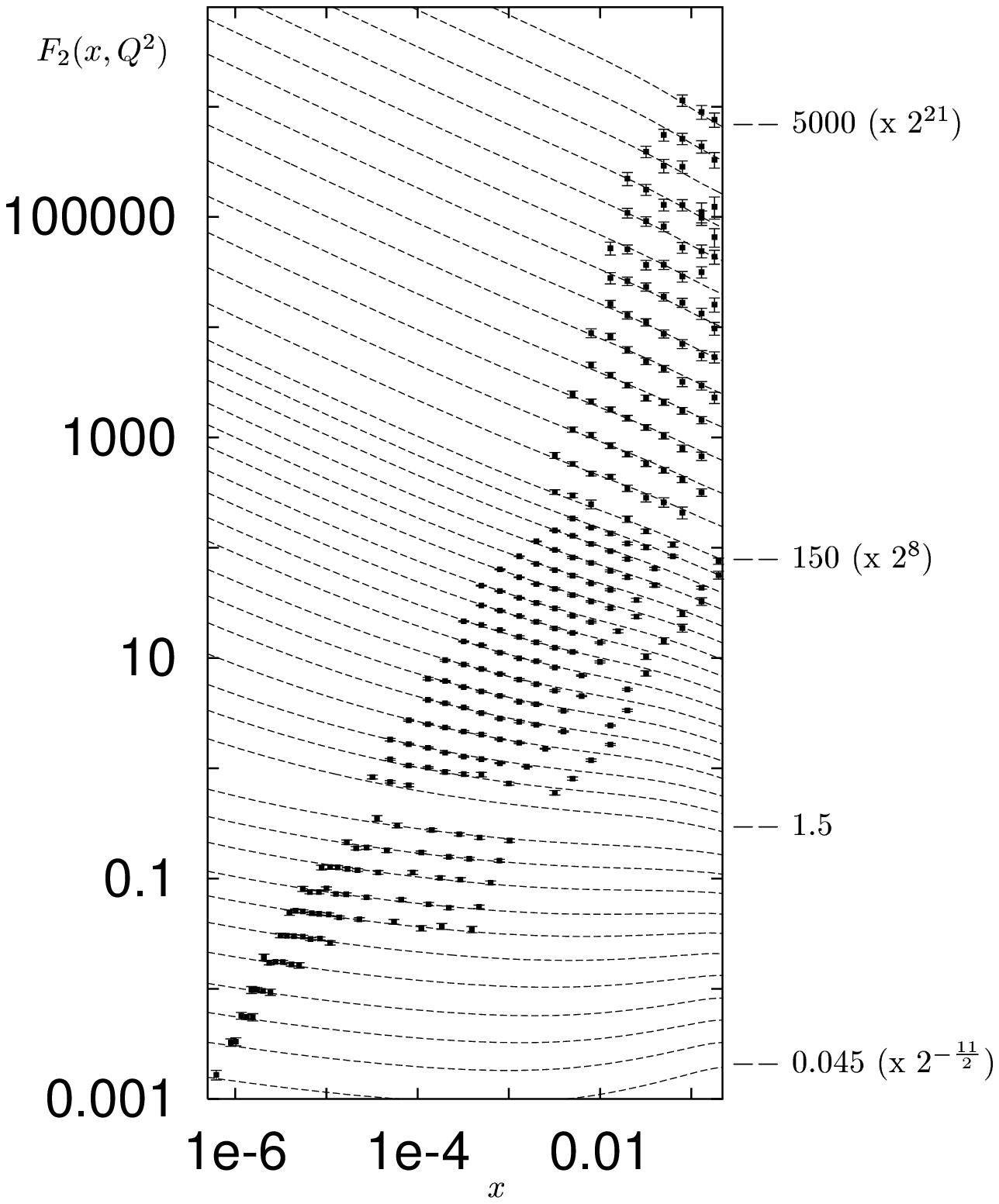}}
\bye

\centerline{\big\Blue{Charm}}

The data for the charm contribution $F_2^c(x,Q^2)$ display a very
interesting simplicity. Even at small $Q^2$, they agree well with a
fixed power of $1/x$ close to 0.4:

\centerline{\epsfxsize=0.52\hsize\epsfbox[100 315 430 750]{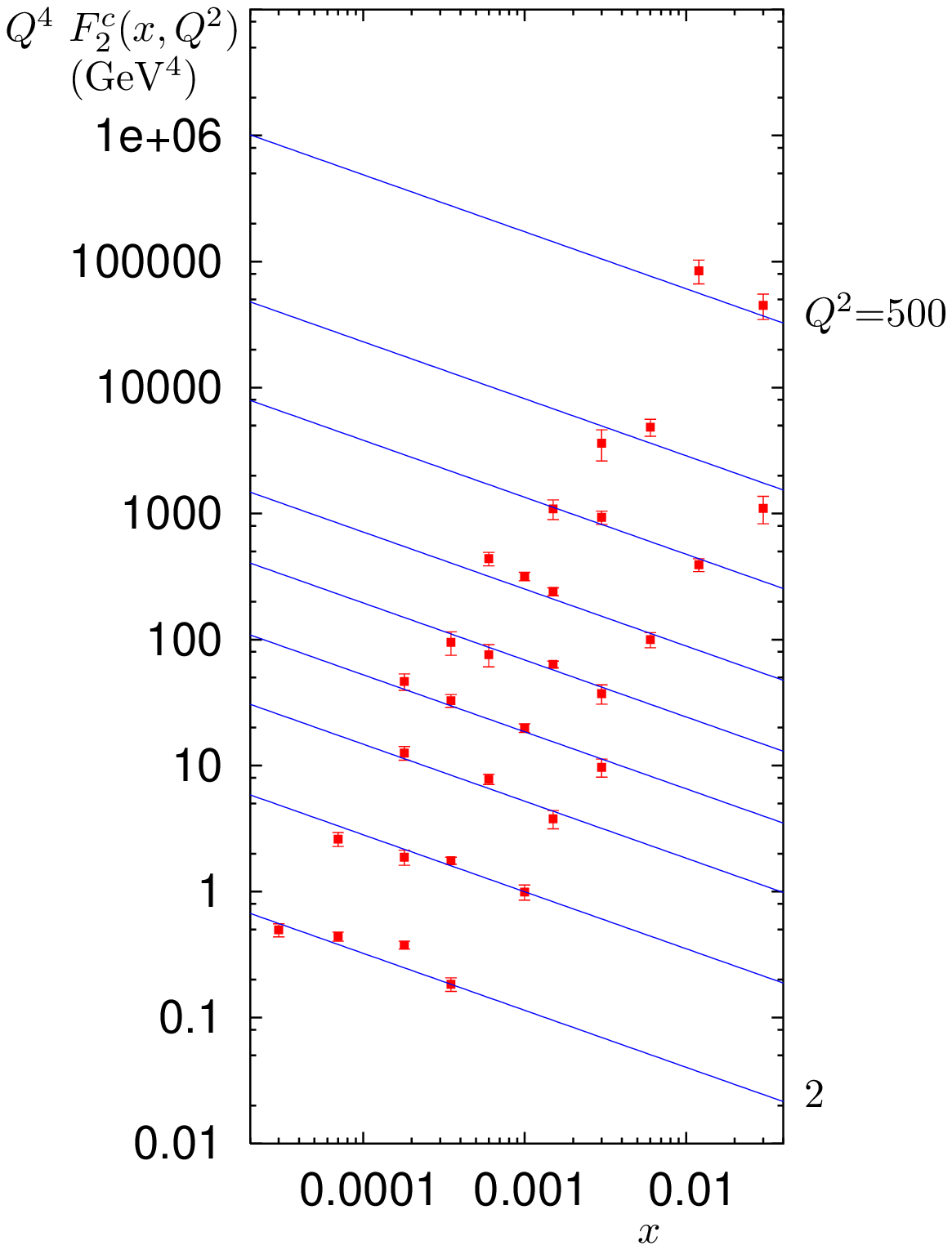}}

The lines are just 2/5 of the hard-pomeron contribution to the
complete $F_2(x,Q^2)$. The factor 2/5 suggests that the coupling of
the hard pomeron to the $c$ quark has the same strength as to the
light quarks. The data do not allow more than an extremely small
contribution from soft-pomeron exchange. 

So only the hard pomeron couples to charm, a result that I find surprising.
It seems to be true even at $Q^2=0$:
\medskip
\centerline{\epsfxsize=0.42\hsize\epsfbox[80 500 300 770]{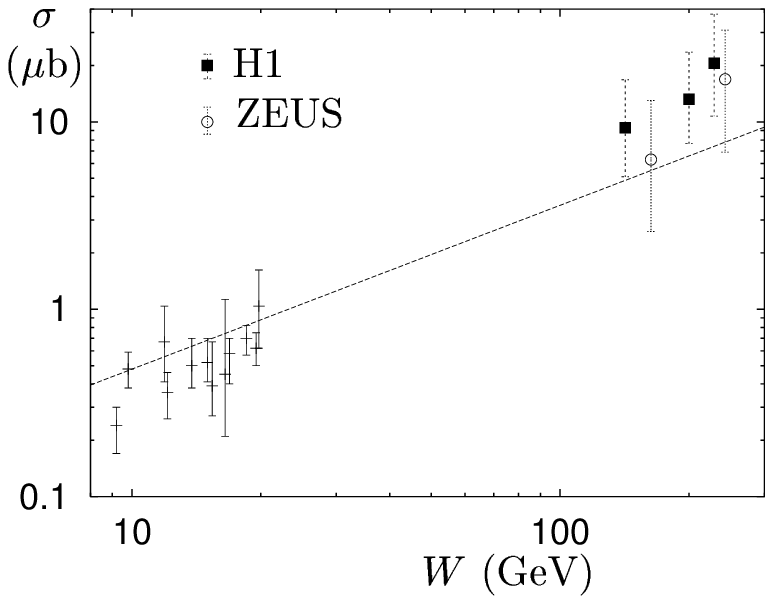}}

\bye

\def\pd{\partial}
\def\u{{\bf u}}
\def\P{{\bf P}}
\def\f{{\bf f}}
\centerline{\big\Blue{Perturbative QCD}}
\bigskip
Perturbative QCD and Regge theory have to live together.

The singlet DGLAP eqation is:
$$
{\pd\over\pd t}\u (x,Q^2)=\int _x^1 dz\, \P (z,\alpha_s(Q^2))
\,\u({x/z},Q^2)
$$$$
\u(x,Q^2)=\left (\matrix{q(x,Q^2)\cr g(x,Q^2)\cr}\right ).
$$
It simplifies if we Mellin transform with respect to $x$.
That is, define
$$
\u(N,Q^2)=\int _0^1dx\,x^{N-1}\u(x,Q^2)
$$$$
\P (N,\alpha_s(Q^2))=\int_0^1 dz\,z^N\P (z,\alpha_s(Q^2)).
$$
Then

$$
{\pd\over\pd t}\u (N,Q^2)= \P (N,\alpha_s(Q^2))\,\u(N,Q^2).
$$

If $\u(x,Q^2)\sim \f(Q^2)x^{-\epsilon}$ at small $x$, then $\u(N,Q^2)$ has 
a pole 
$$
\f(Q^2)\over N-\epsilon
$$
Insert this in the  DGLAP equation. The pole singularities on the two sides of the equation must balance as $N\to\epsilon_0$ :
$$
\Blue{{\pd\over\pd t}\f(Q^2)= \P (N=\epsilon,\alpha_s(Q^2))\, \f(Q^2).}
$$
For hard-pomeron exchange, with $\epsilon\approx 0.4$, expand the matrix
$\P(\epsilon,\alpha_s)$ in powers of $\alpha_s$.
This is \BrickRed{\underbar{not}} valid for soft-pomeron exchange,
$\epsilon\approx 0.08$, because the elements of $\P$ are singular at $N=0$.

\b This is a problem with all applications of pQCD to the evolution of 
$F_2(x,Q^2)$, though usually it is a problem that is hidden.

\bye

To solve the differential equation, we need initial conditions at some
$Q^2$, eg $Q^2=20$ Gev$^2$. 
We saw that for $F_2^c(x,Q^2)$ soft-pomeron exchange is extremely small.
So pQCD suggests that this is true also for $g(x,Q^2)$.
\bigskip
{\epsfxsize=0.45\hsize\epsfbox[150 520 470 770]{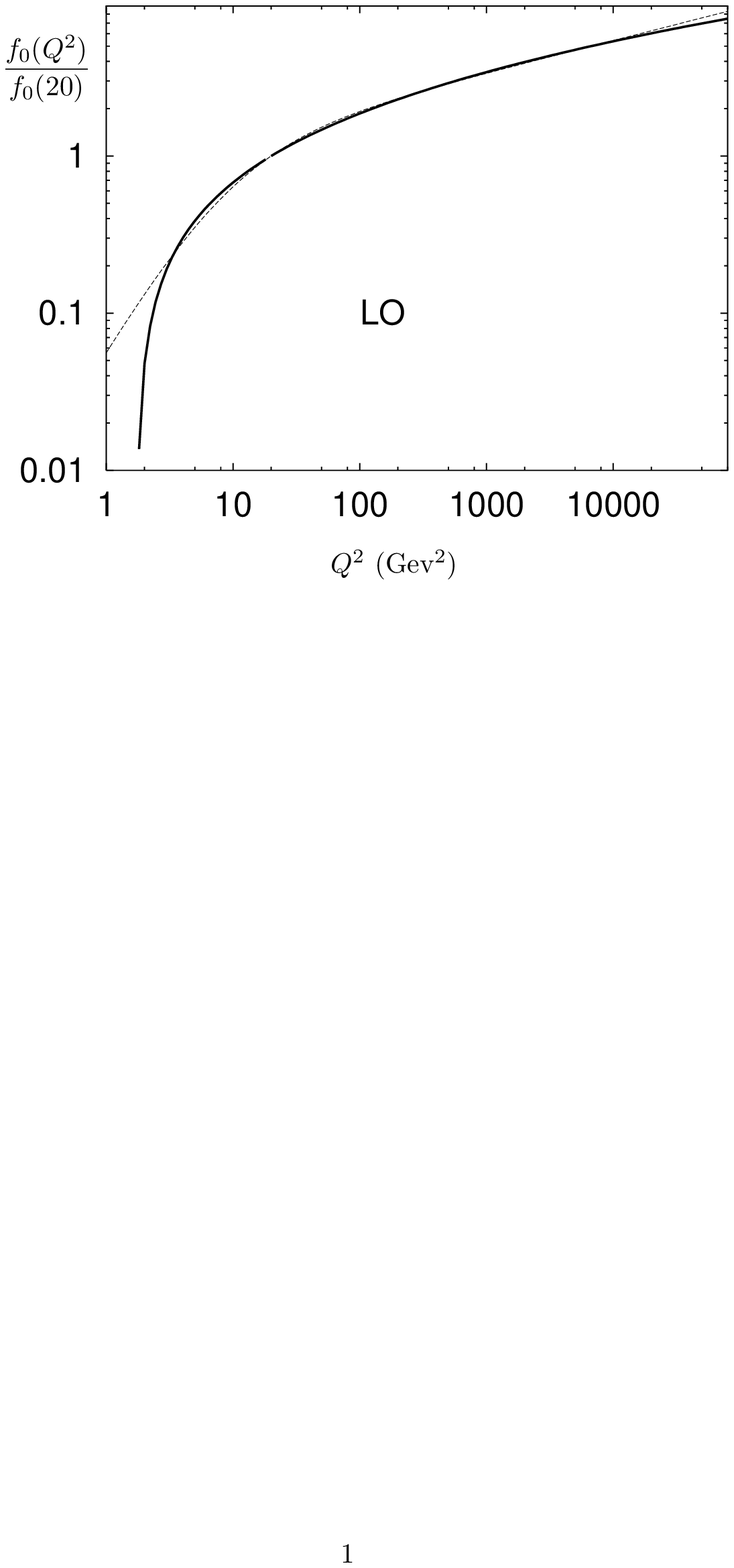}\hfill
\epsfxsize=0.45\hsize\epsfbox[150 520 470 770]{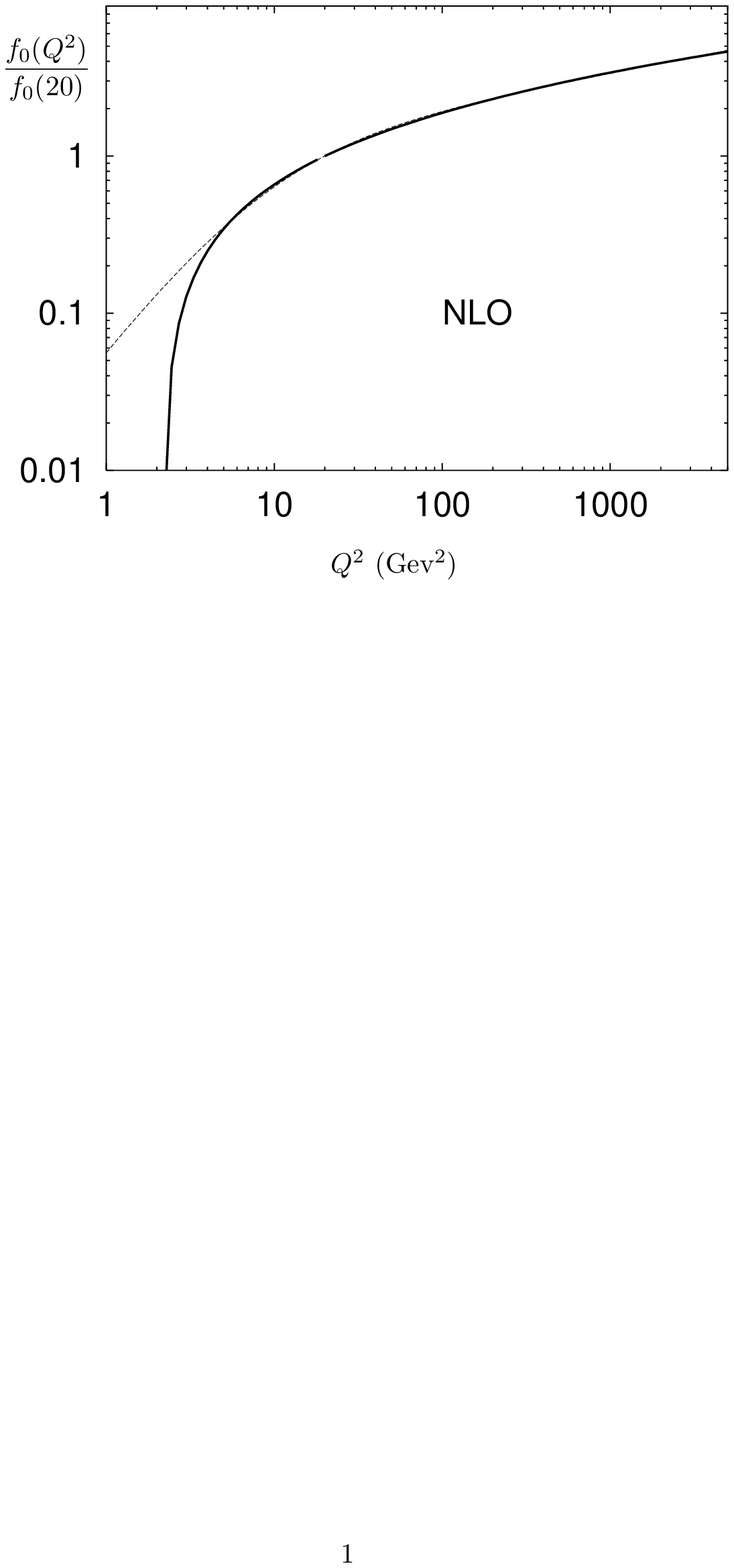}}
\bigskip

The upper curves come from the fit to the data that I have described.
The lower curves are the solution to the DGLAP equation, using values
of $\Lambda_{\hbox{{\sevenrm QCD}}}$ similar to what are usually
accepted:
$$
\Lambda^{\hbox{\rm LO}}=140~\hbox{MeV}~~~~~~~~~ 
\Lambda^{\hbox{\rm NLO}}=330~\hbox{MeV}
$$
Note that the DGLAP equation is supposed to be valid only for large $Q^2$;
evidently this means $Q^2$ greater than about 5~GeV$^2$. It is interesting
that the LO and NLO results are not very different. 

It is interesting also that the upper curves behave as a power of $Q^2$
at large $Q^2$, because that is how we chose to parametrise $f_0(Q^2)$.
But the outputs from the DGLAP calculations rather behave as power
of log$(Q^2)$. Numerically, they are almost identical over a very large
range of $Q^2$ values.
\bye

Because the gluon density is dominated at all $Q^2$ by hard pomeron
exchange alone, that is it behaves approximately as $x^{-0.4}$ for all
$Q^2$, it is rather larger than is conventionally supposed,
particularly at small $Q^2$:
\bigskip
{\epsfxsize=0.48\hsize\epsfbox[50 50 385 285]{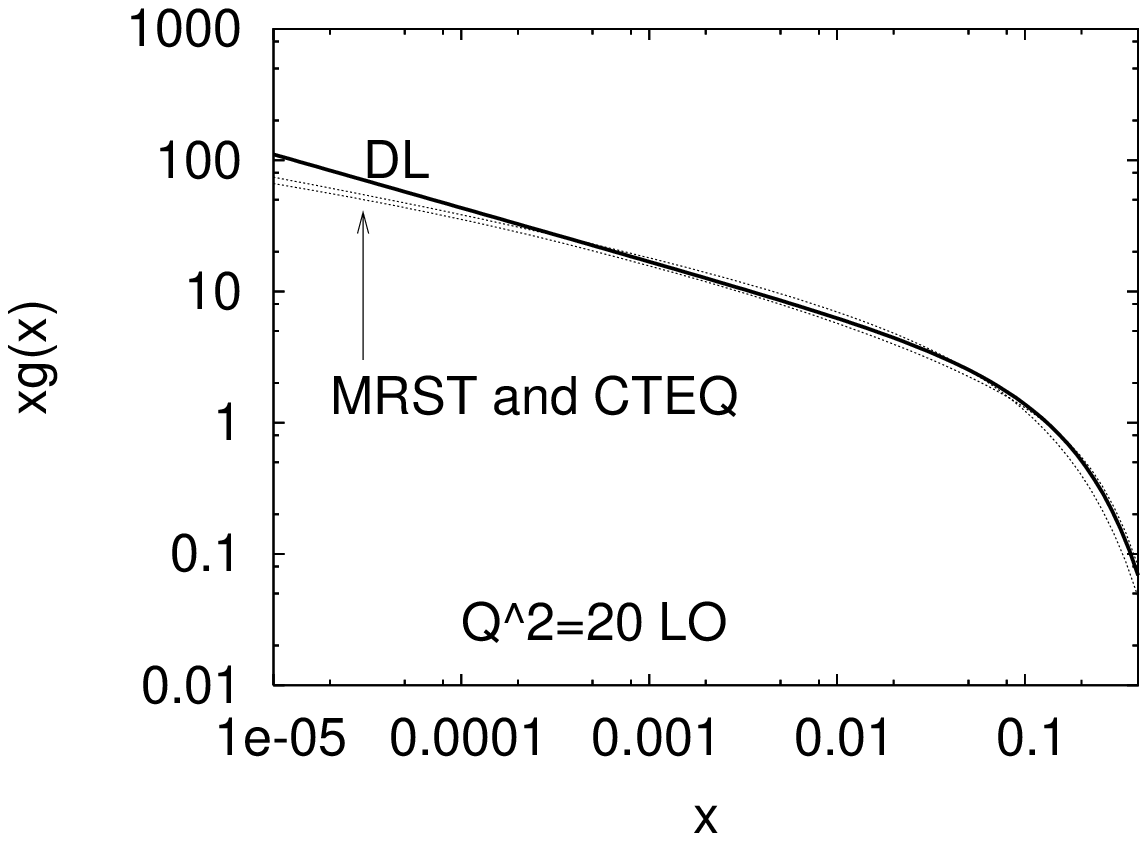}\hfill
\epsfxsize=0.48\hsize\epsfbox[50 50 385 285]{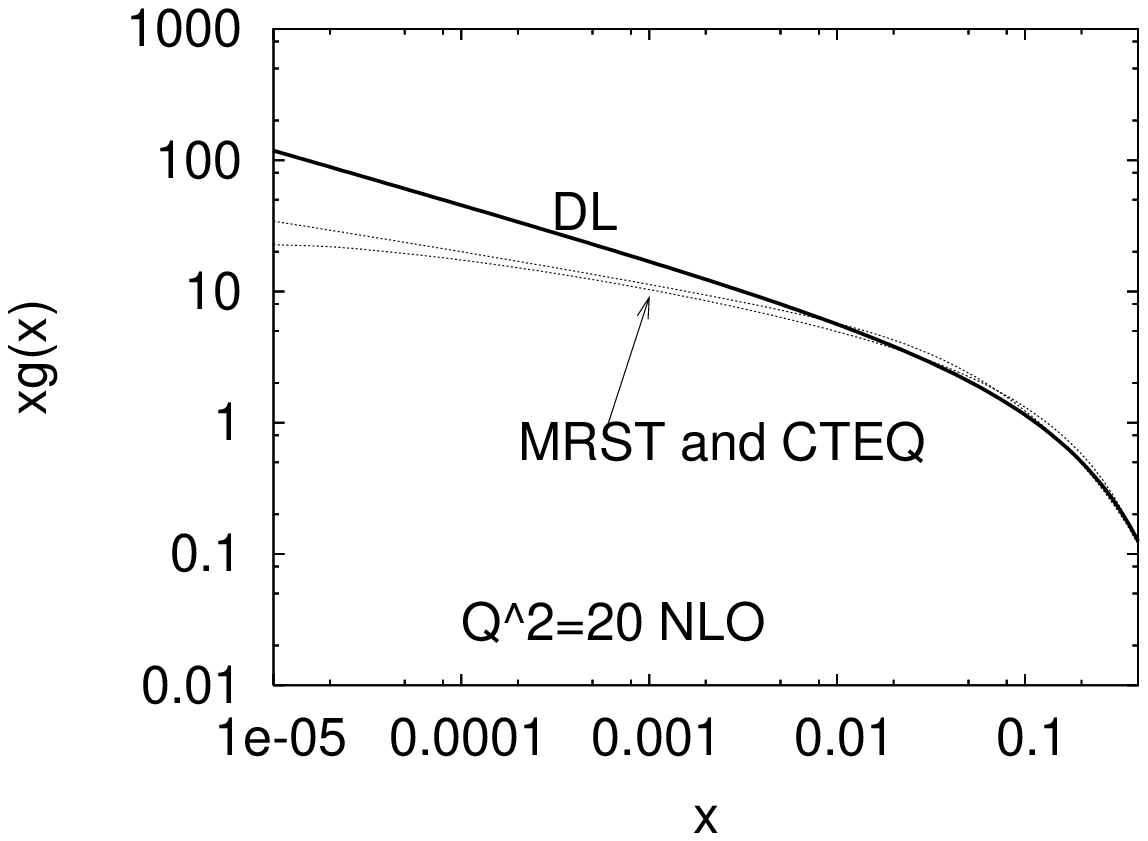}}
\vskip 10truemm
\bye

\centerline{\bf\Blue{Regge factorisation}}
\vskip 8truemm
For each of the separate exchanges  hardpom, softpom and reggeon
\bigskip
\centerline{\epsfxsize=0.5\hsize\epsfbox{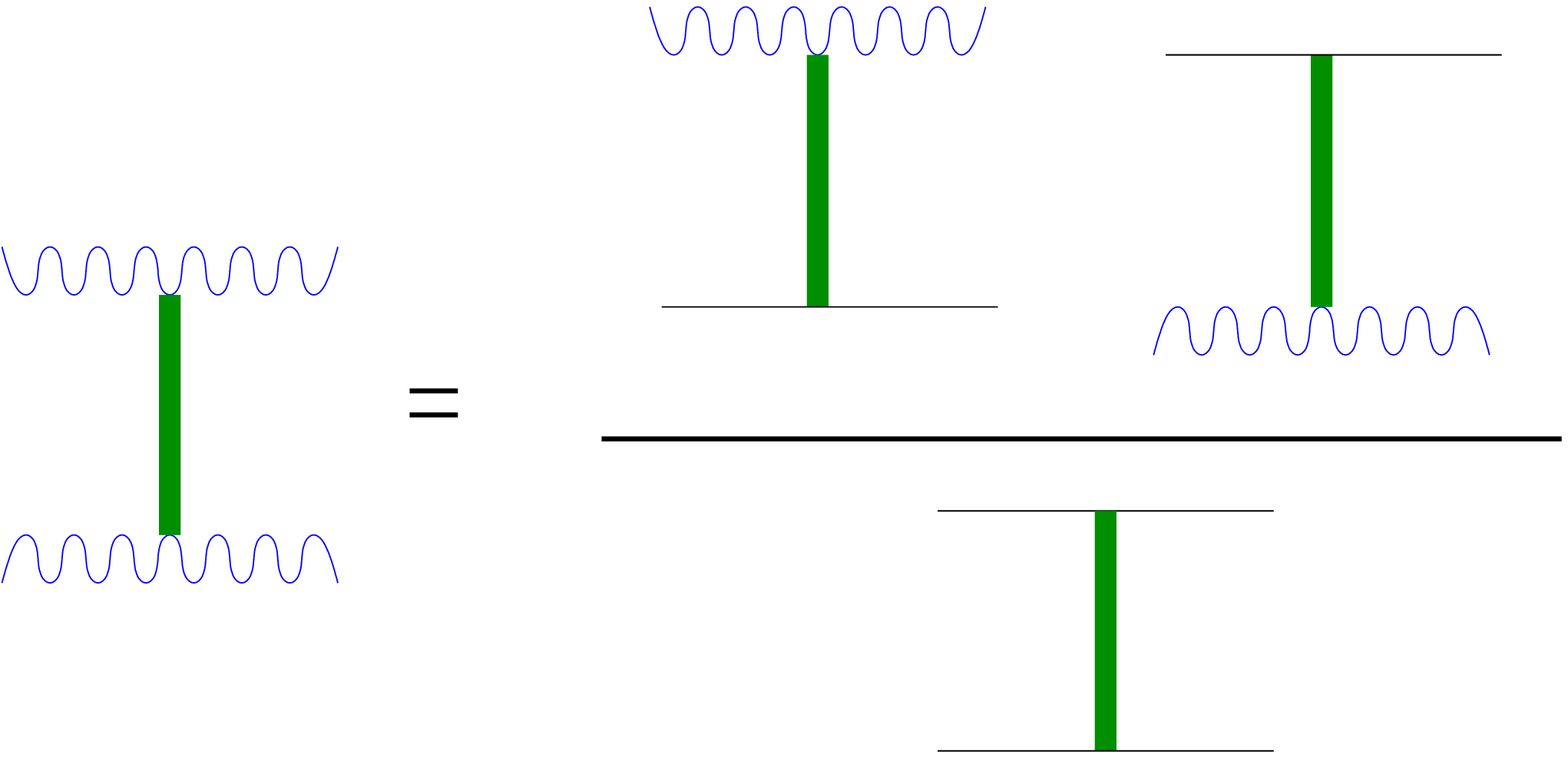}}
\bigskip
This is valid if the exchange corresponds to a pole in the
complex-$\ell$ plane, because for each exchange the contribution
is the product of a coupling at each vertex and a ``propagator'' corresponding
to the other factors in the Regge formula, including the power of $s$.
Thus, for each term, the contribution satisfies
$$
\sigma(\gamma\gamma)={\sigma(\gamma p)\sigma(\gamma p)\over\sigma(pp)}
\hbox{ for all }Q_1^2,Q_2^2~~~~~~~~~~~
$$

For $\gamma\gamma$ we must add in the box graph
\bigskip
\centerline{\epsfxsize=55truemm\epsfbox{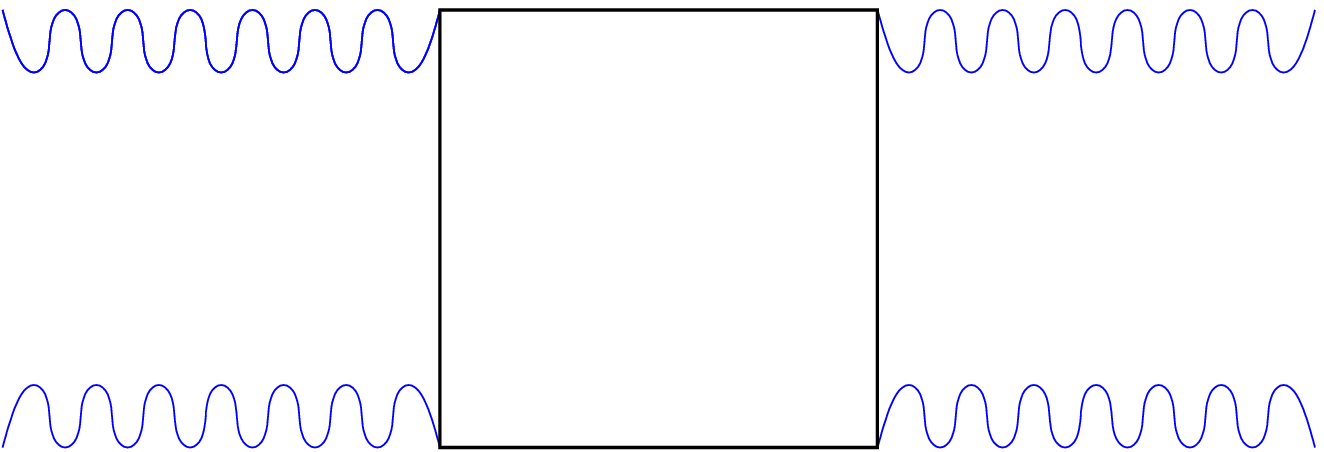}}
\bigskip
summed over the possible quark flavours in the loop. It is particularly
important when the energy is not very large.
\bye

In the case of real photons, the fits I have described for
$\sigma(\gamma p)$ and $\sigma(pp)$ yield
for $\sigma(\gamma\gamma )$
\bigskip
\centerline{\epsfxsize=0.44\hsize\epsfbox[110 500 485 775]{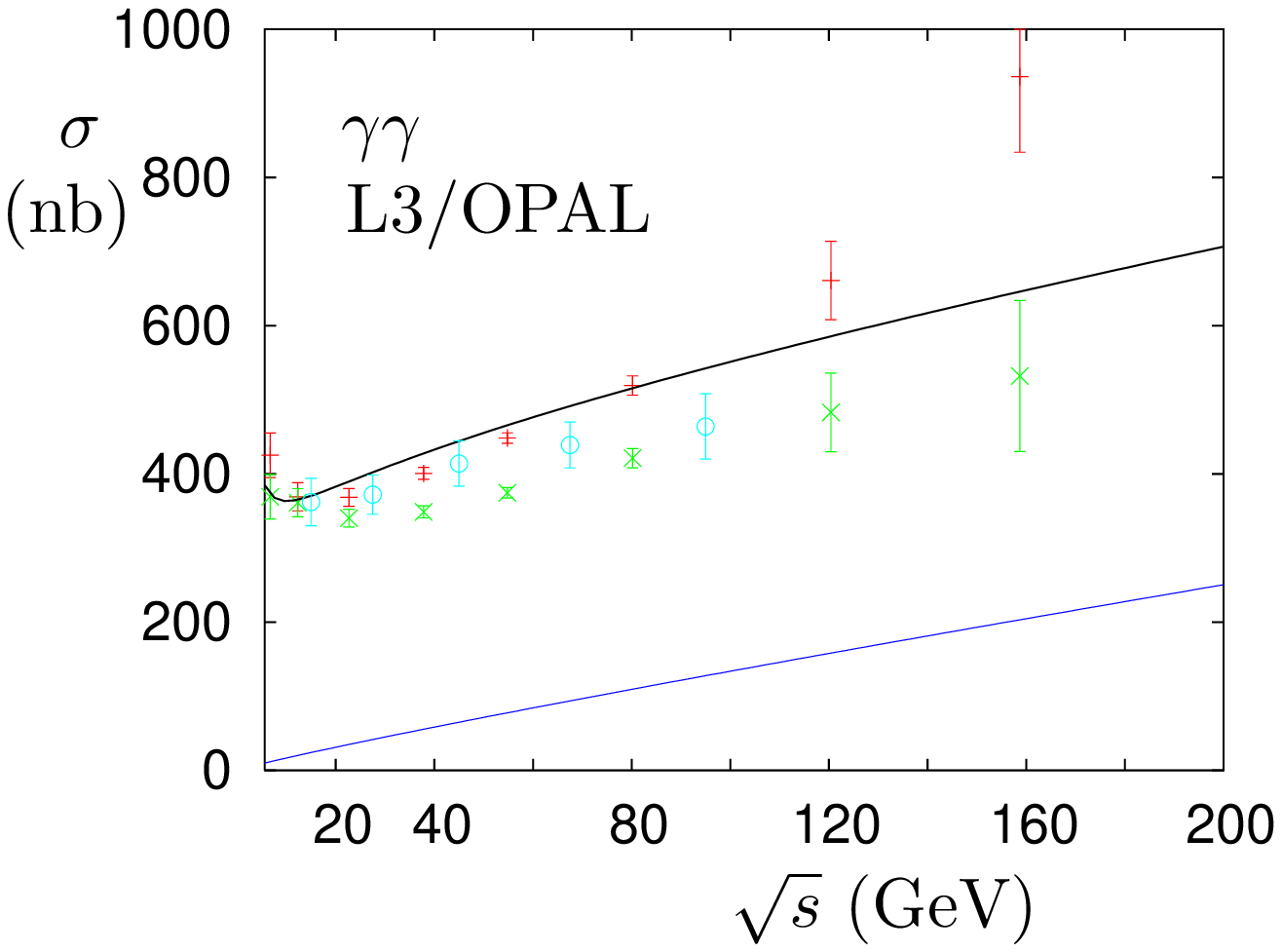}}
\bigskip
The experimental data are uncertain because of the need to make
large acceptance corrections, and different models for calculating these
lead to rather different outputs, as is seen in the plot.

We may similarly calculate the charm component of $\sigma(\gamma\gamma )$:
\bigskip
\centerline{\epsfxsize=0.44\hsize\epsfbox[110 500 485 775]{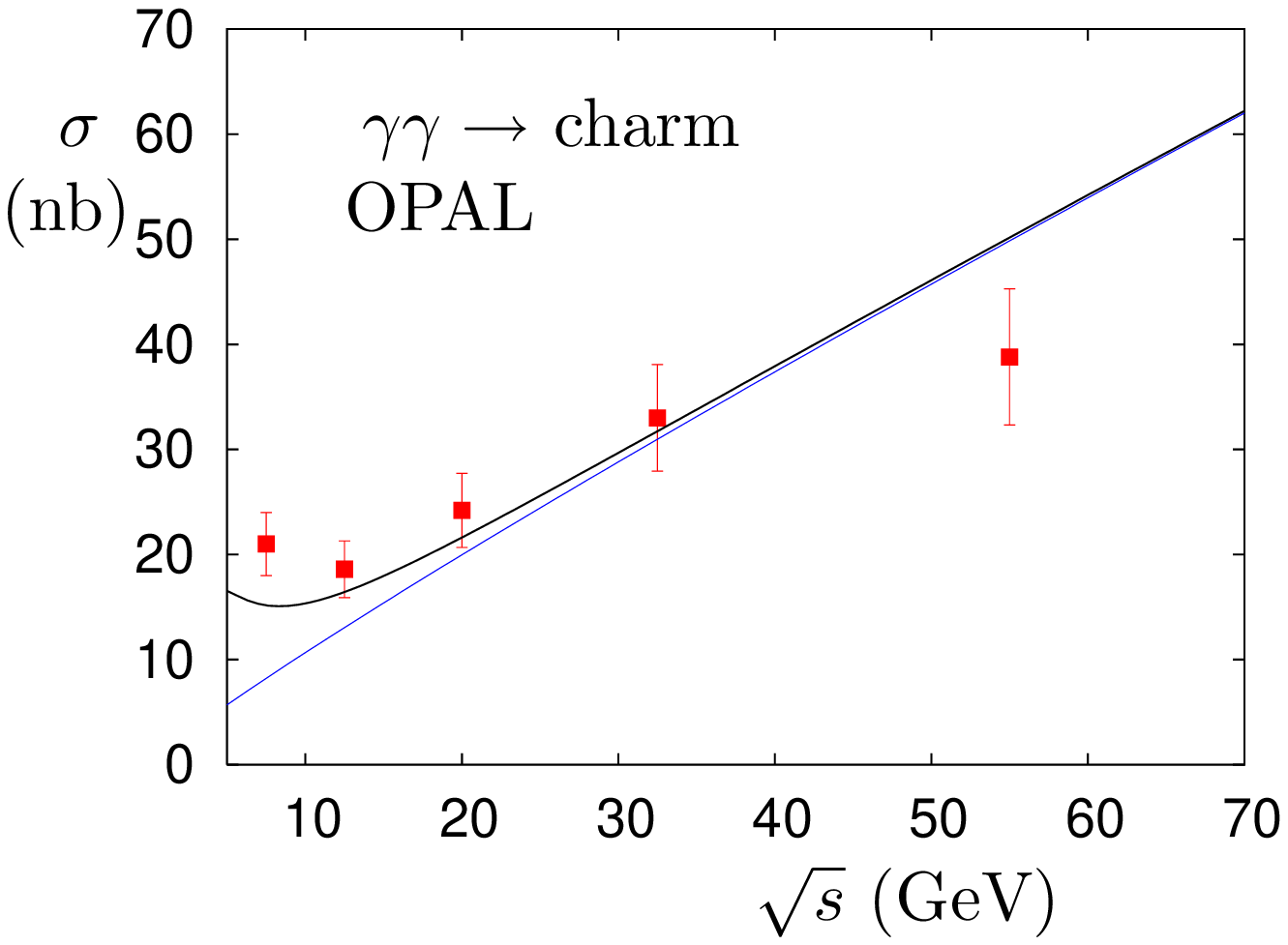}}
\vskip 10truemm
In each case, the blue curves are the hard-pomeron contribution. The 
box graph is what causes the cross-section initially to fall with increasing
$W$.
\bye

\centerline{\big\Blue{Photon structure function}}
\bigskip
When one photon is on shell and the other off shell, a similar calculation
gives
\bigskip
\centerline{\epsfxsize=0.6\hsize\epsfbox[130 500 470 755]{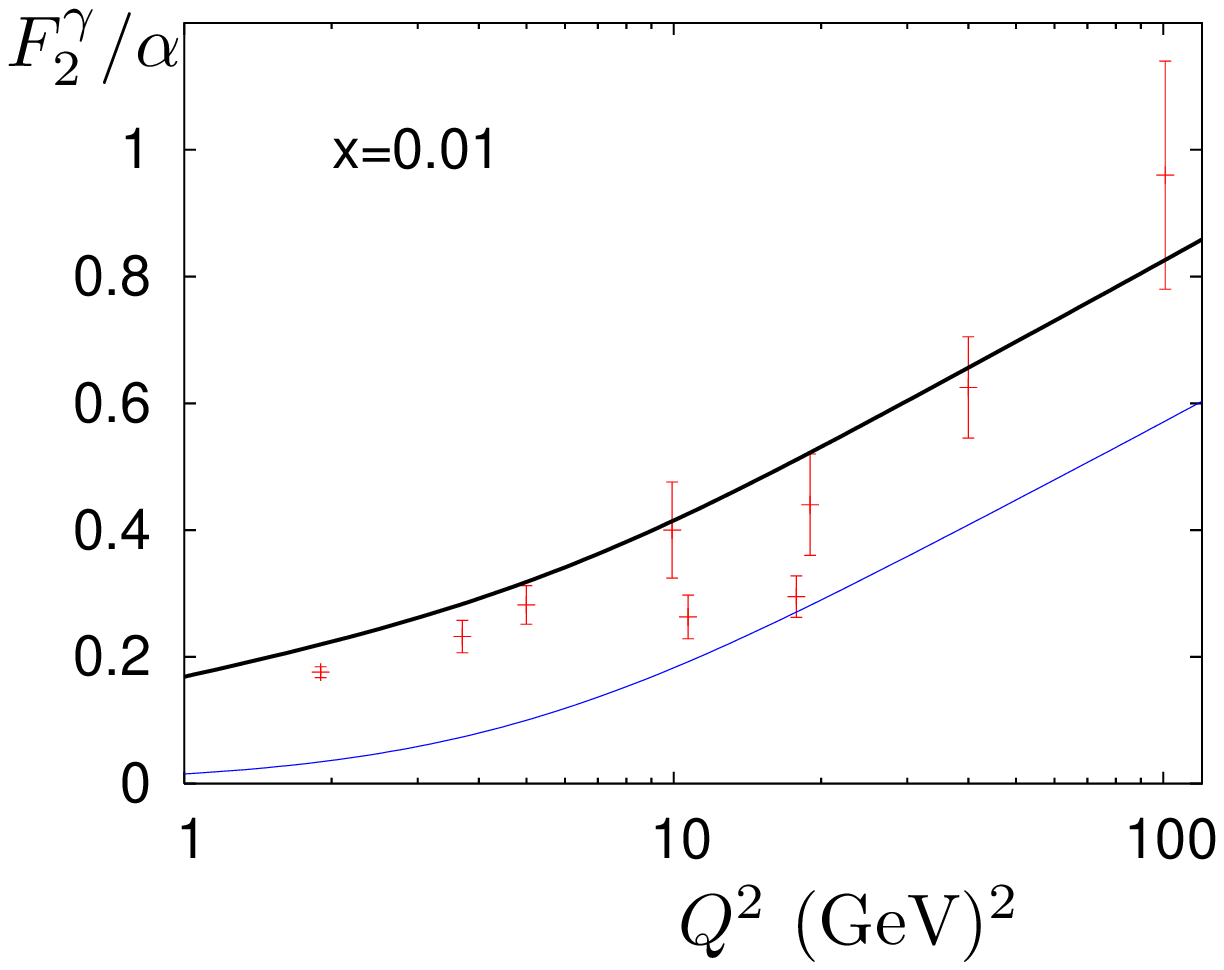}}
\bigskip

and when both photons are off shell
\bigskip
\centerline{\epsfxsize=0.7\hsize\epsfbox[50 500 470 770]{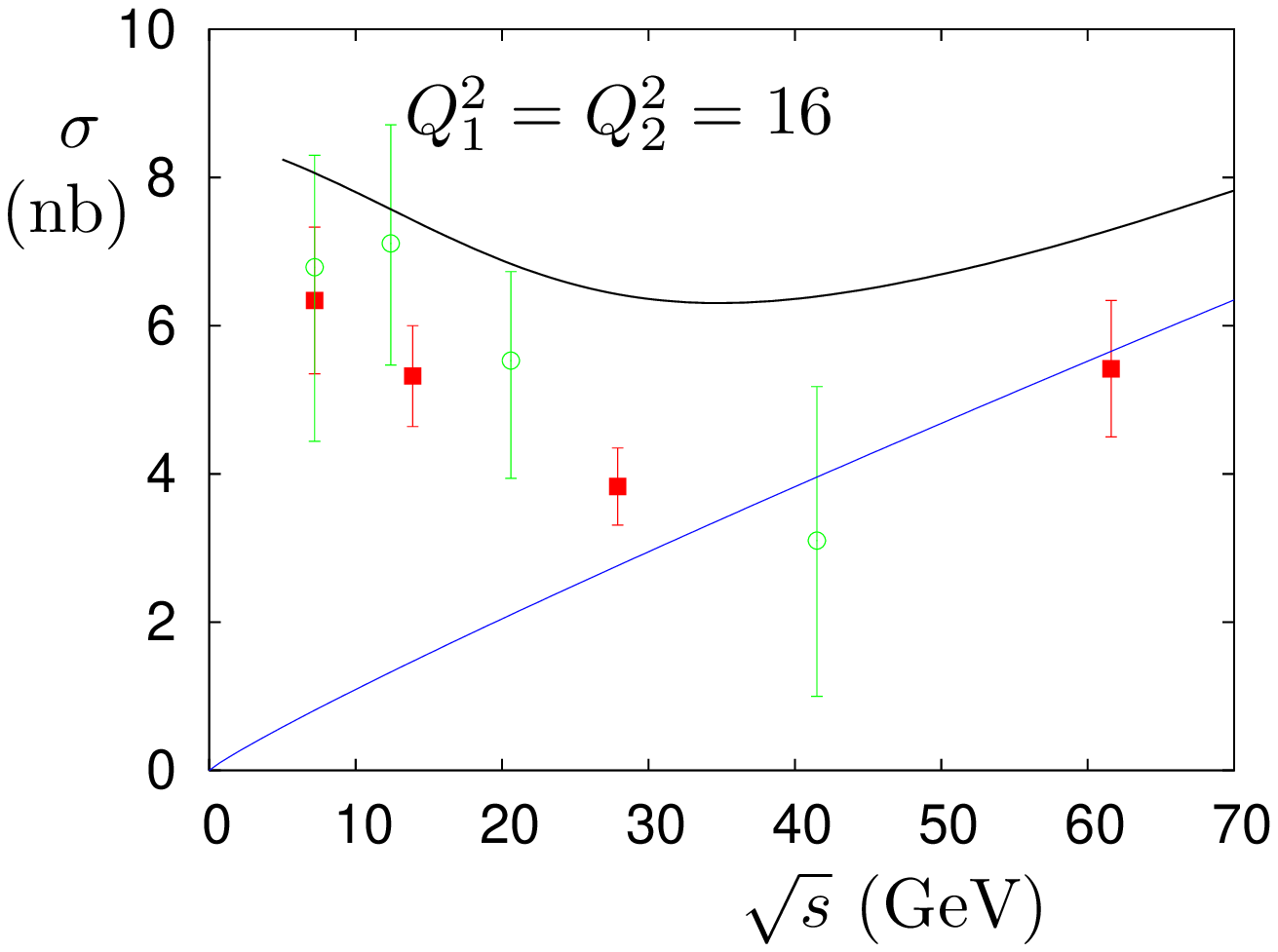}\hfill}
\vskip 20truemm
\BrickRed{Conclusion about Regge factorisation:}\h
\centerline{I'm not sure, the data are not good enough.}
\bye

\centerline{{\big \Blue{$\gamma p\to J/\psi ~p$}}}
\bigskip
Parametrise the amplitude as a sum of 
hardpom + softpom + reggeon exchange and adjust the relative
contributions so as to fit the data for $d\sigma/dt$. 
Because the Regge signature factor gives each term a different
phase, the resulting amplitude is not altogether simple.
the result is
\bigskip
\centerline{\epsfxsize=0.55\hsize\epsfbox[55 560 390 780]{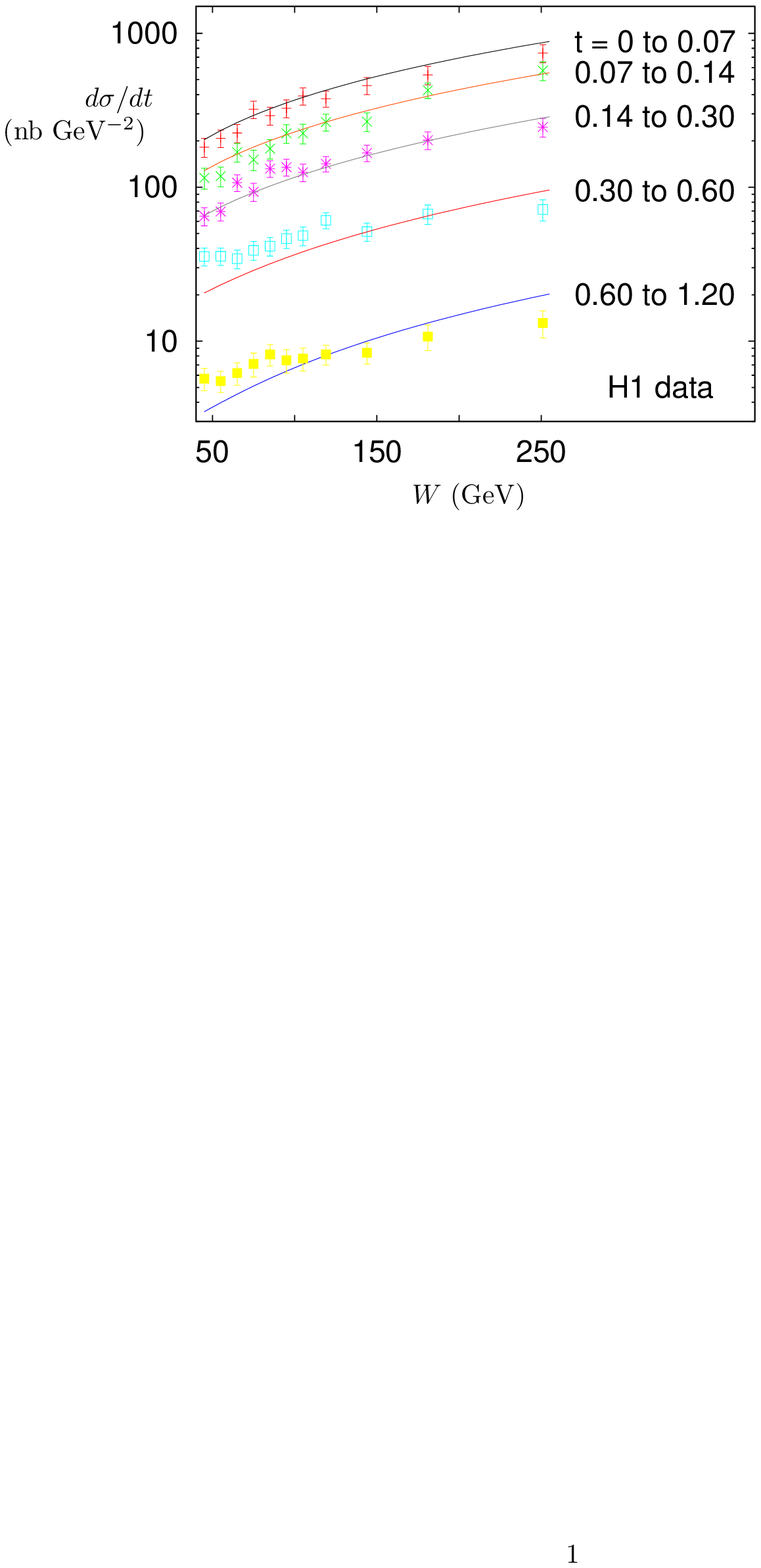}}
\bigskip
The fit needs the hard pomeron slope to be quite small, perhaps 0.05.
Integrating over $t$, we find
\bigskip
\centerline{\epsfxsize=0.55\hsize\epsfbox[55 560 390 780]{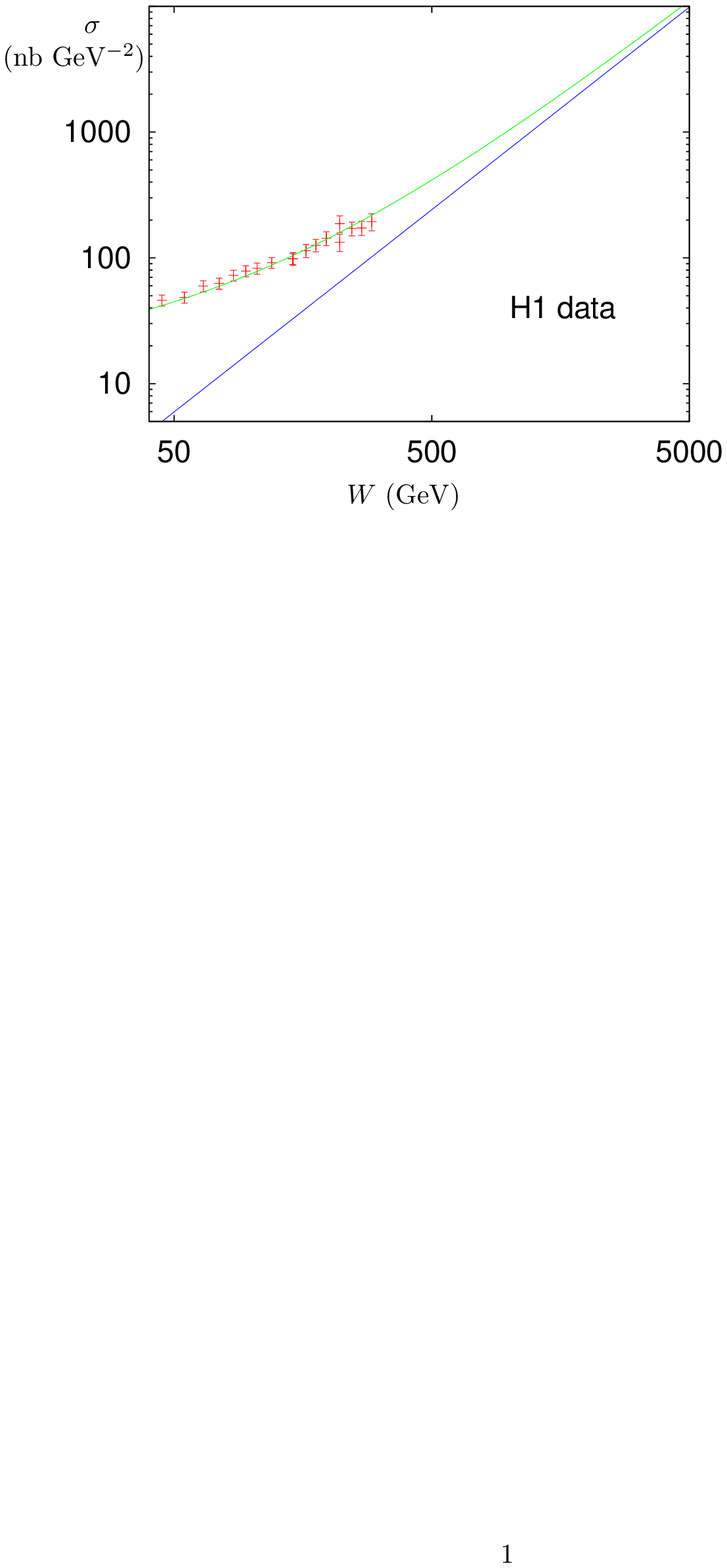}}
\bigskip
The blue line is from the hard-pomeron term alone.
Why does the soft pomeron couple to $\gamma p\to J/\psi\, p$ but not to
$F_2^c$? The explanation may be found in data from the
Omega experiment (Physics Letters 68B (1977) 96):
$$
{\bar p\,\hbox{Cu}\to J/\psi\, X \over p\,\hbox{Cu}\to J/\psi\, X}\approx 6
$$
So the valence quarks of the beam couple to $J/\psi$:
{\Blue{$|c\bar c\rangle$ mixes with $|q\bar q>$}} and the $J/\psi$
is not pure $c\bar c$.

\bye

\centerline{\Blue{{\big Hard pomeron in hadron-hadron scattering?}}} 
\vskip 8truemm
Given that the hard pomeron exists, does it contribute to $pp$ and
$p\bar p$ scattering? Try including its contribution in the fits
to the total cross sections. 
That is, use hardpom, softpom and reggeon exchange:
\bigskip
$\sigma(pp),\sigma(p\bar p),\sigma(\gamma p)$:
$
~~~~~\sigma=X_0s^{\epsilon_0}+X_1s^{\epsilon_1}+X_Rs^{\epsilon_R}
$
\bigskip
$F_2(x,Q^2)$:
$
~~~~~~~~~~~~~~~x^{-\epsilon_0}f_0(Q^2)+x^{-\epsilon_1}f_1(Q^2)+x^{-\epsilon_R}f_R(Q^2)
$
\bigskip
The resulting fits, with
$$\epsilon_0=0.45~~~~~~~~\epsilon_1=0.067~~~~~~~~\epsilon_R=-0.48$$
are:
\bigskip

\line{$\phantom{X}$\epsfxsize=0.45\hsize\epsfbox[125 495 470 770]{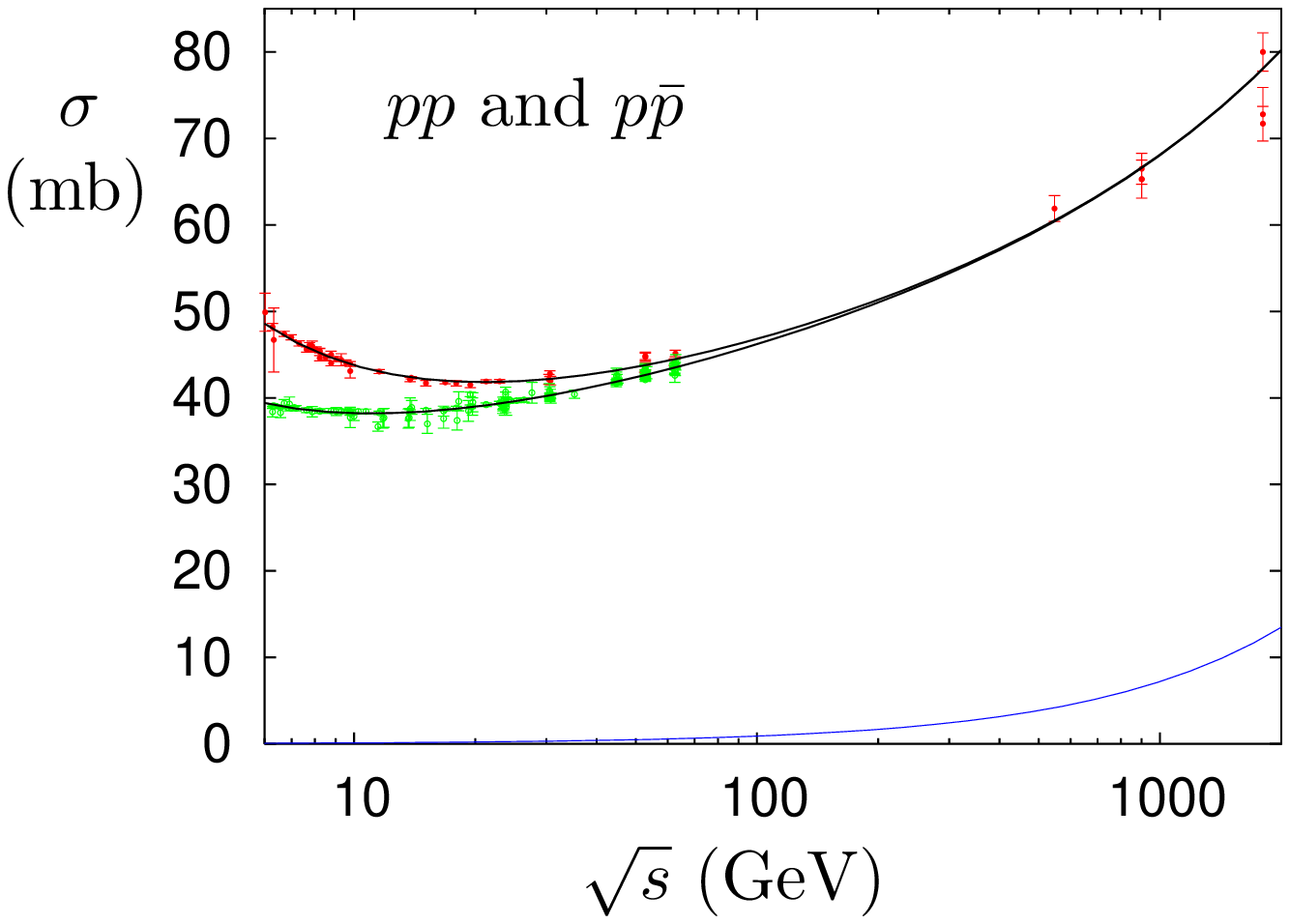}
\hfill
\epsfxsize=0.45\hsize\epsfbox[125 495 470 770]{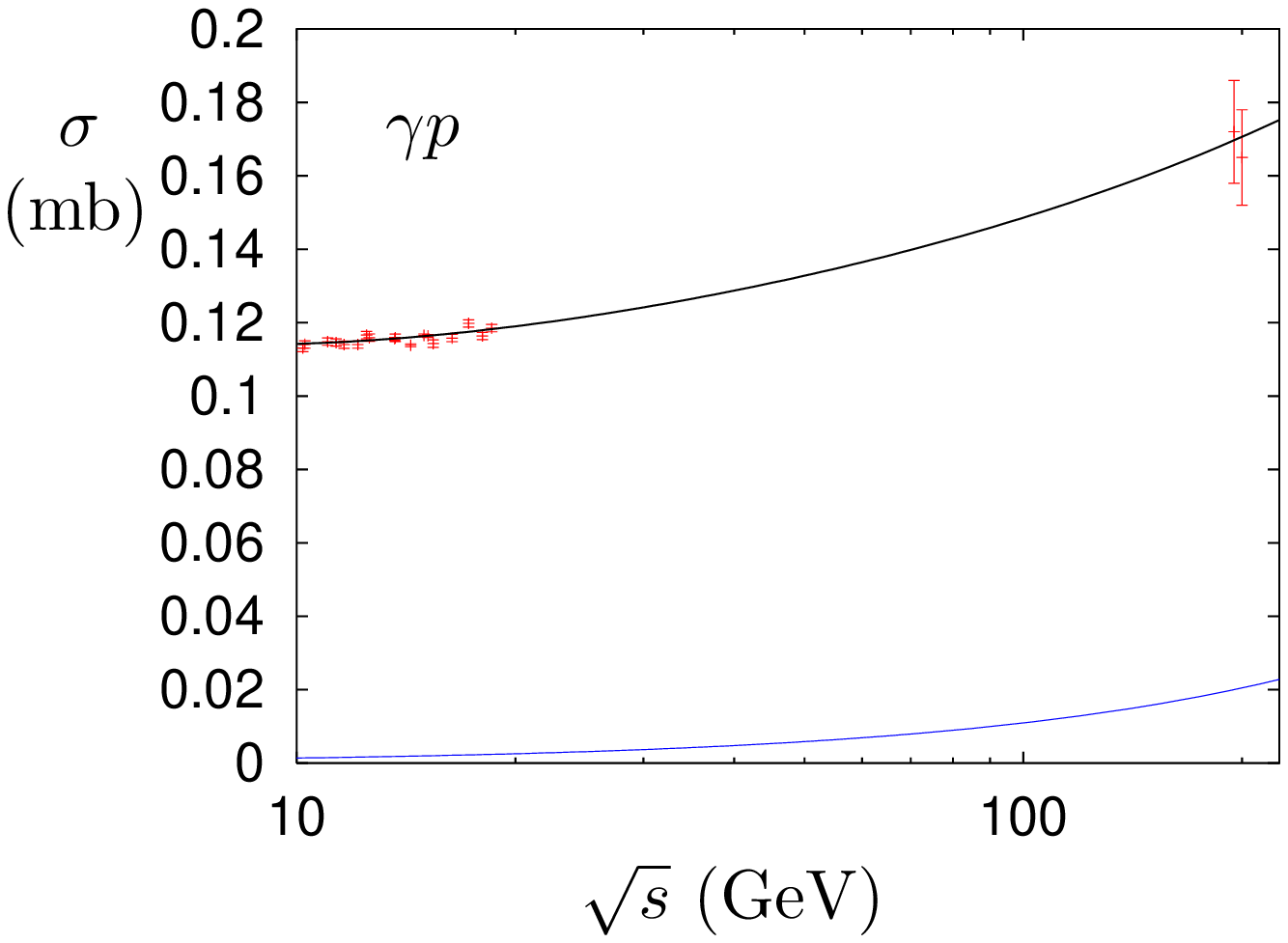}}
\bigskip
The blue lines are the hardpom contribution.
\bye

\centerline{\bf \Blue{Extrapolate to LHC energy}}
\bigskip
\centerline{\epsfxsize=0.45\hsize\epsfbox[125 495 470 770]{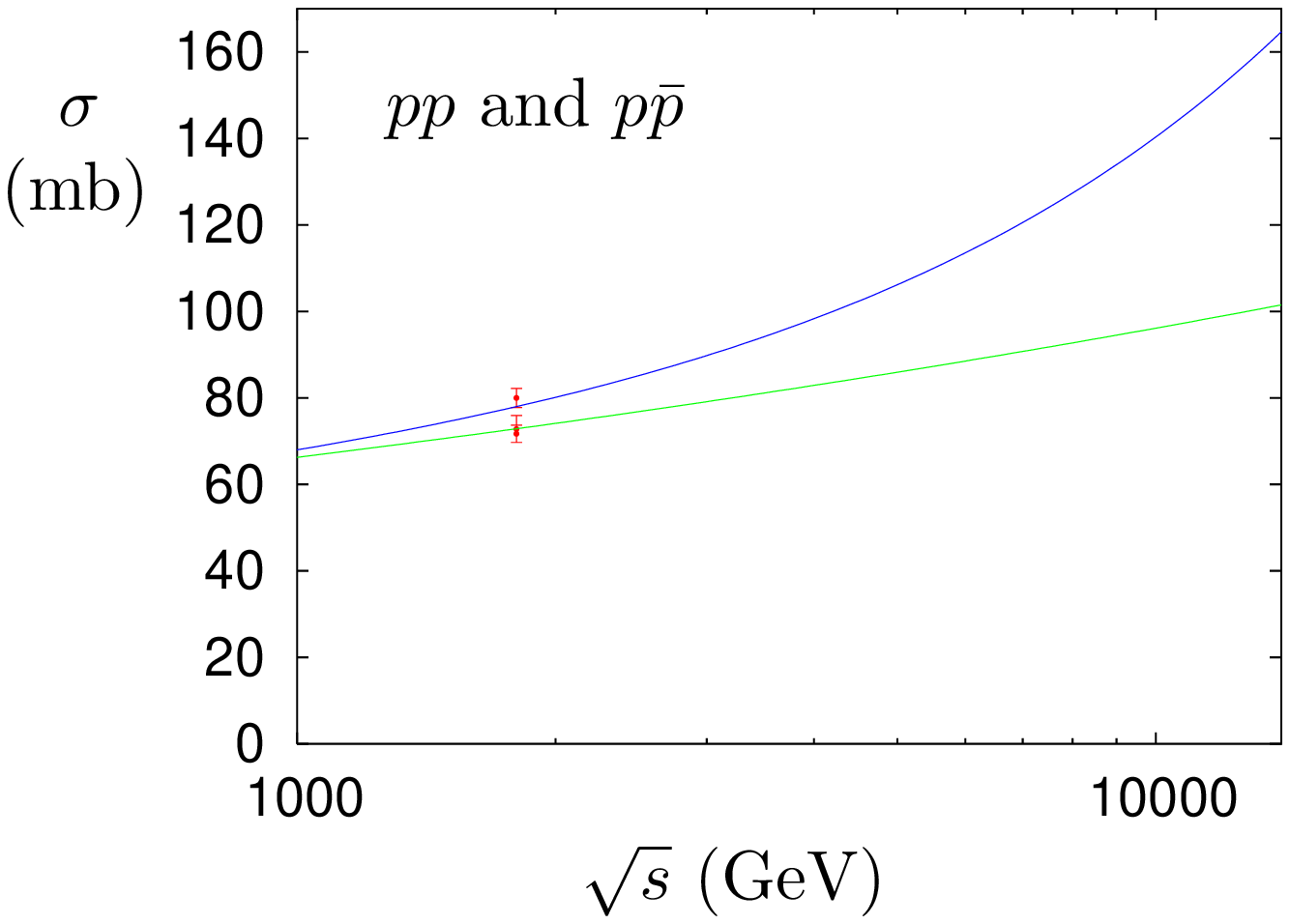}}
\bigskip
The lower line is the original fit with a hardpom contribution.

We have to worry about unitarity.

\b \Blue{We do not know how to do that!}

\def\P{I\!\!P}
We need to sum single-$\P$, double-$\P$, \dots exchanges:
\bigskip
\epsfxsize \hsize\epsfbox{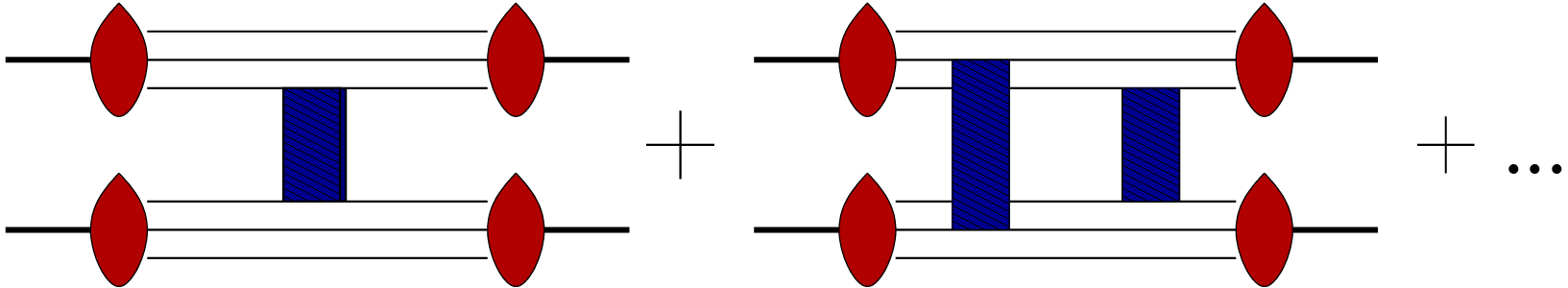}
\bigskip
Although we know some general properties of these additional terms, we
cannot calculate them. The double-$\P$-exchange term has the general structure
$$
A_{\P _1\P _2}(s,t) \sim \beta (t)~s^{\alpha_{\P _1\P _2}(t)}(\log(s))^{-\gamma(t)}
$$
For linear trajectories,
$$
\alpha_{\P _1\P _2}(t)=\alpha_{\P _1\P _2}(0)+\alpha_{\P _1\P _2}^{\prime}t
$$$$
\eqalign{
\alpha_{\P _1\P _2}(0)&=\alpha_{\P _1(0)}+\alpha_{\P _2}(0)-1\cr
\alpha_{\P _1\P _2}^{\prime}&={\alpha'_{\P _1}\alpha'_{\P _2}\over \alpha'_{\P _1}+\alpha'_{\P _2}}\cr
}$$
But $\beta (t)$ and $\gamma(t)$ are unknown. Evidently, $\beta (t)$ depends
on information about two-quark correlations in the proton wave function.
\bye

\def \p{{\bf p}}
\def \q{{\bf q}}
\centerline{\Blue{{\big Eikonal model for} {\bigit pp} {\big scattering}}}
\bigskip
This model has
\underbar{no} theoretical foundation, but it produces multiple-exchange terms of the correct general structure.
\bigskip
In the CM frame
$$
\left . \matrix{
p_1=(E,\p +\half \q )&p_3=(E,\p -\half \q )\cr
p_2=(E,-\p -\half \q )&p_4=(E,-\p  + \half \q)\cr
}\right .
$$
with $(\p +\half \q )^2=(\p -\half \q ) ^2$ so that $\p .\q =0$  and therefore
$\q $ is in the two-dimensional space perpendicular to $\p$. Also $t=-\q ^2$.

Write the amplitude as a 2-dimensional Fourier integral
$$\eqalign{
A(s, -{\bf q}^2)&=4 \int d^2b\, e^{-i{\bf q}.{\bf b}} \tilde{A}(s,{\bf b}^2)\cr
\tilde{A}(s,{\bf b}^2)&={1\over 16 \pi^2}\int d^2q\, e^{i{\bf q}.{\bf b}}A(s, -{\bf q}^2)\cr
}$$

${\bf b}$ is called the \underbar{impact parameter}.
\bigskip
Define
$$
\chi(s,b)=-\log(1+2i\tilde A/s)
$$
so that
$$
\tilde{A}(s,{\bf b}^2)= \half is\big(1-e^{-\chi(s,b)}\big)
$$
Remember the unitarity condition

$$
\hbox{Im }a_\ell (s)=|a_\ell (s)|^2 ~+~\hbox{inelastic terms}
$$
so that
$~~~~~~~~~~~~~~~~~~~
|a_\ell (s)|<1
$
\bigskip
One can show that this is satisfied if
$$
\hbox{Re }\chi(s,b)\ge 0
$$
\bye

Expand the exponential as a power series:
$$
\eqalign{
A(s, -{\bf q}^2)
&=2is\int d^2b\, e^{-i{\bf q}.{\bf b}}
\big(1-e^{-\chi(s,b)}\big)\cr
&=2is\int d^2b\, e^{-i{\bf q}.{\bf b}} ~\Big(\chi-{{\chi^2}\over{2!}}+
{{\chi^3}\over{3!}} \dots -{{(-\chi)^n}\over{n!}}\dots\Big)\cr
}
$$
If first term is approximated by single-$\P$ exchange, the second has the 
correct general structure of
double-$\P$ exchange, etc.
And one can then show that then at very large $s$
$$
\sigma^{\rm Tot}\sim 4\pi\alpha'\epsilon_0 (\log s)^2
$$
so the Froissart bound is satisfied.
\bigskip
{\big \BrickRed{\underbar{But}}}, although it has been widely used, this representation for the amplitude has little theoretical foundation. 
For example, the double-exchange term should contain information about
the two-quark correlation in the proton's wave function, but this is
not present in the term $\chi^2$. 
\bye

\centerline{\Blue{{\bigit pp} {\big elastic scattering}}}
\bigskip
The fit to $d\sigma/dt$ using just the two single pomeron exchanges
and reggeon exchange agrees well with the data at small $t$, but not 
at larger $t$:
\bigskip
\centerline{\epsfxsize=0.5\hsize\epsfbox[50 50 410 300]{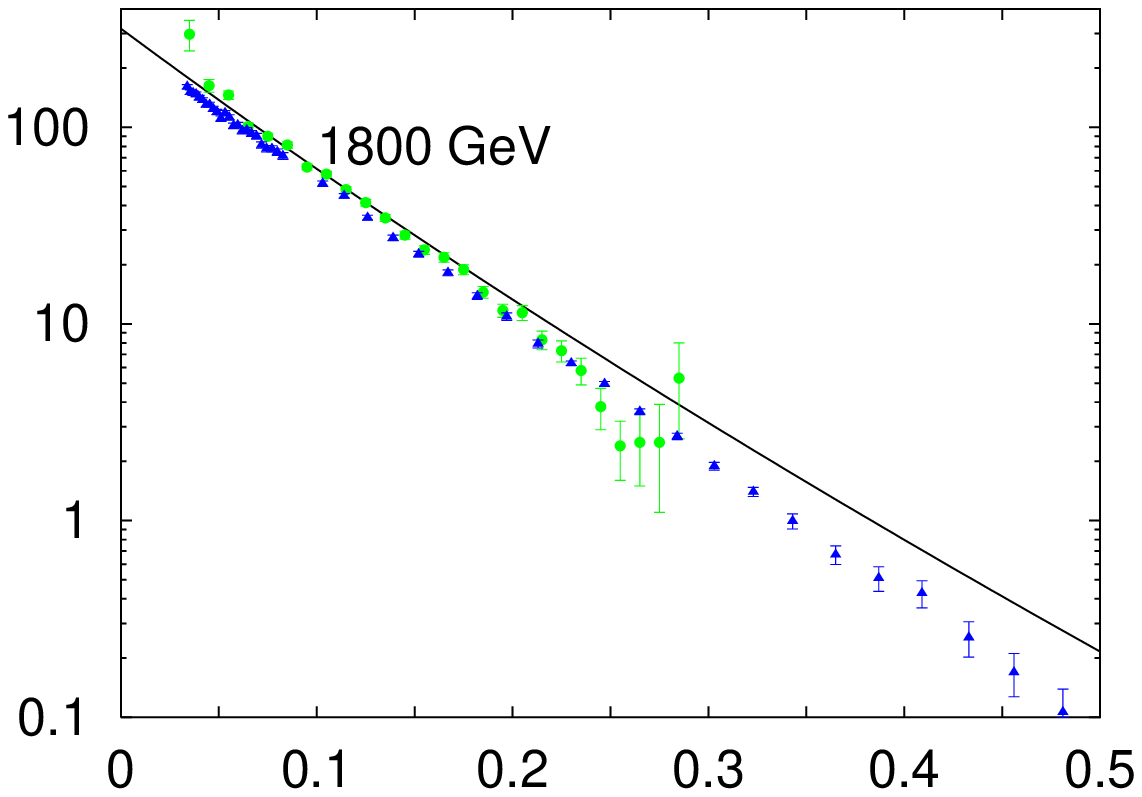}}
\bigskip
From its general structure, we know that
$\P\P$ exchange pulls  $d\sigma/dt$ down at larger $t$.

\Blue{But nobody knows how to calculate it!}

As a simple model, calculate $\chi(s,b)$ as the sum of the three
single exchanges and take 
$$\tilde A(s,b)=2is\Big (\chi(s,b)-\lambda [\chi(s,b)]^2\Big )$$
together with a triple-gluon exchange term.
Then repeat the fit, choosing $\lambda$ to get $pp$ dips at the right $t$:
\bigskip

\centerline{\epsfxsize=0.95\hsize\epsfbox[80 560 530 770]{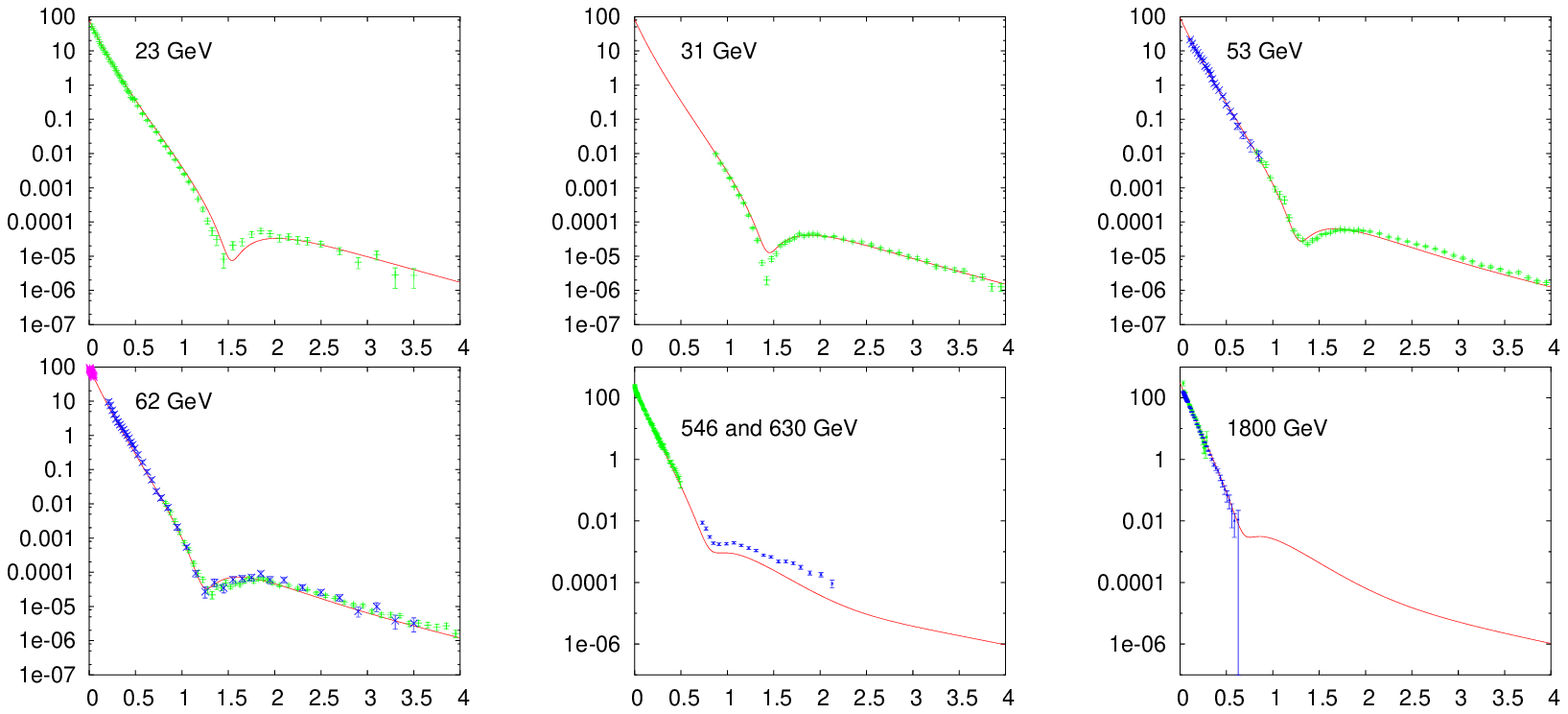}}
\bye

\centerline{\big\Blue{\underbar{Conclusion}}}
\bigskip
The extrapolation to LHC energy is now the upper curve in
\bigskip
\centerline{\epsfxsize=0.6\hsize\epsfbox[50 50 385 285]{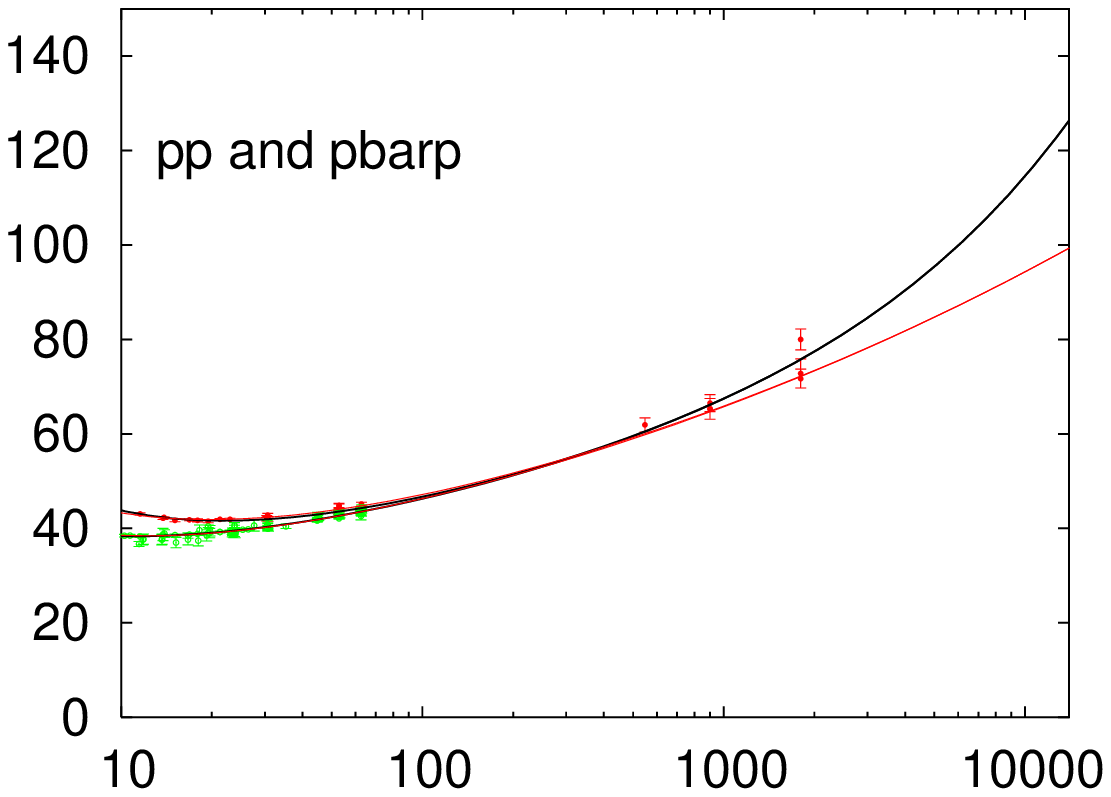}}
\bigskip
The lower curve is the fit with no hard-pomeron contribution.
\vskip 25truemm

This procedure is highly model-dependent, so the error is inevitably
large:

\b{{$\sigma(\hbox{LHC})=125\pm 25~\hbox{mb}$}}
\bye

\end